\documentclass[11pt]{article}

\usepackage[final]{acl}
\usepackage{times}
\usepackage{latexsym}
\usepackage[T1]{fontenc}
\usepackage[utf8]{inputenc}
\usepackage{microtype}
\usepackage{graphicx}
\usepackage{multirow}
\usepackage{booktabs}
\usepackage{makecell}
\usepackage{pifont}
\usepackage{subcaption}
\usepackage{amsmath}
\usepackage{amssymb}
\newcommand{\cmark}{\ding{51}}
\newcommand{\xmark}{\ding{55}}

\title{Beyond Retrieval: Learning Compact User Representations for Scalable LLM Personalization}

\author{
    Heng Cao$^*$\thanks{This work was done while Heng Cao was at Microsoft.}, Fan Zhang$^*$, Jian Yao$^\dagger$\thanks{Corresponding author.}, Yujie Zheng$^\diamond$, Changlin Zhao$^*$, Lu Hao$^*$, \\
    \bfseries Yuxuan Wei$^*$, Wangze Ni$^\ddagger$, Huaiyu Fu$^*$, Yuqian Sun$^*$, Xuyan Mo$^*$ \\
    \normalfont $^*$Microsoft \quad $^\diamond$Shanghai International Studies University \quad $^\ddagger$Zhejiang University \\
    $^\dagger$The Hong Kong Polytechnic University \\
    \texttt{hengcao2026@gmail.com} \quad \texttt{nigel97.yao@connect.polyu.hk} \\
    \texttt{\{zhanfa, neilz, luhao, yuxwei,} \\
    \texttt{huaiyufu, yuqiansun, xuyanmo\}@microsoft.com} \\
    \texttt{zhengyujie@shisu.edu.cn} \quad \texttt{niwangze@zju.edu.cn}
}

\begin{document}
\setcounter{footnote}{3}
\begin{NoHyper}
\maketitle
\end{NoHyper}

\begin{abstract}
Personalizing large language models requires adapting model behavior to individual users while preserving robustness and deployment-scale efficiency.
Existing approaches typically personalize LLMs either at the input level, by retrieving user histories or constructing profile prompts, or at the parameter level, by maintaining user-specific parameter-efficient modules.
The former makes personalization sensitive to retrieval quality and prompt design, whereas the latter incurs storage and maintenance costs that grow with the user population.
To address these limitations, we propose \textbf{TAP-PER}\footnote{Code: \url{https://github.com/heng380/TAP-PER}.} (\textbf{T}emporal \textbf{A}ttentive \textbf{P}refix for \textbf{PER}sonalization), a prefix-based framework that encodes user preferences as learnable representations, avoiding the serialization of user histories into prompts and replacing heavy per-user adapters with lightweight user-state prefix embeddings.
Inspired by personalized recommendation systems, TAP-PER decomposes user modeling into user-state and query-conditioned components, and incorporates temporal signals to capture the evolving nature of user interests.
Experiments on six LaMP tasks show that TAP-PER consistently outperforms prompt-based and model-based baselines across classification, rating, and generation settings.
Moreover, TAP-PER uses $130\times$ fewer per-user parameters than OPPU and roughly half the total parameter footprint of PER-PCS at the 1,000-user scale, demonstrating scalable personalization without prompt-serialized histories or heavy per-user adapters.
\end{abstract}

\section{Introduction}
Large language models (LLMs)~\cite{brown2020gpt3,grattafiori2024llama,yang2025qwen3} have achieved strong performance across a wide range of natural language processing tasks, yet their outputs often remain generic: the same input is typically answered in the same way regardless of the user's long-term interests, stylistic preferences, or task-specific goals. This limitation has motivated growing interest in personalized LLMs~\cite{liu2025survey,chen2024when,kirk2024benefits}, which aim to adapt model behavior to individual users while retaining the general capabilities of pretrained language models.

\begin{figure}[t]
  \centering
  \includegraphics[width=\columnwidth]{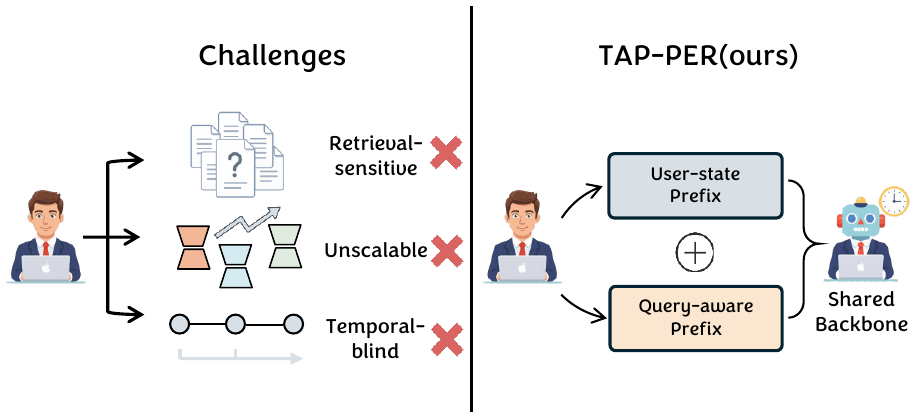}
  \caption{Existing personalized LLMs face three challenges: retrieval sensitivity, poor scalability, and temporal blindness (left). TAP-PER addresses them via a user-state prefix and a temporal-aware query-conditioned prefix injected into a shared backbone (right).}
  \label{fig:challenge}
\end{figure}

Existing approaches to LLM personalization largely fall into two categories. The first category personalizes LLMs at the input level. Retrieval-augmented generation (RAG) methods retrieve relevant user histories and prepend them to the query, while profile-augmented generation (PAG) methods summarize user histories into natural-language profiles~\cite{chen2024when,li2023teach,mysore-etal-2024-pearl,richardson2023integrating}. These methods are flexible and require no parameter updates, but personalization is delegated to how user context is retrieved, summarized, and serialized into the prompt.
The second category personalizes LLMs at the parameter level, typically through parameter-efficient fine-tuning (PEFT). For example, One PEFT Per User (OPPU) trains a dedicated adapter for each user~\cite{tan2024democratizing}, while PER-PCS reduces redundancy by decomposing and recombining personalized adapter components~\cite{tan2024personalized}.
More recent methods use progressive LoRA-based mixture-of-experts or hypernetworks to generate personalized adapters from user profiles~\cite{zhang2025proper,tan2025instant}. These methods provide a parameterized route to personalization by encoding user preferences directly in model parameters.
Despite recent progress, existing personalized LLM methods still face three practical limitations.

\paragraph{Retrieval sensitivity.}
Prompt-based methods such as RAG and PAG are sensitive to retrieval and prompt quality. Table~\ref{tab:main} shows that the optimal RAG/PAG setting varies across tasks, suggesting limited robustness and possible privacy risks from exposing user histories.

\paragraph{Limited scalability.}
PEFT-based methods often allocate dedicated parameters (e.g., LoRA modules) for individual users, resulting in a parameter footprint that grows with the user population. Representative approaches such as OPPU~\cite{tan2024democratizing} and Profile-to-PEFT (P2P)~\cite{tan2025instant} either maintain user-specific parameters or generate them on-the-fly via a hypernetwork. PER-PCS~\cite{tan2024personalized} reduces redundancy through collaborative decomposition, but still requires a sizable pool of personalized parameters. OPPU and PER-PCS are also commonly combined with optional RAG/PAG variants, while P2P consumes a textual user profile to generate an adapter, adding another user-context pipeline.

\paragraph{Temporal unawareness.}
Most methods treat histories as static context and overlook preference evolution, although recent interactions are often more indicative of current intent.

These limitations lead us to ask: \emph{Can we achieve LLM personalization without serializing user histories into prompts or maintaining heavy per-user adapters, while remaining scalable to large populations?}

Inspired by personalized recommendation systems, which model user preferences as learned embeddings, we propose \textbf{TAP-PER} (\textbf{T}emporal \textbf{A}ttentive \textbf{P}refix for \textbf{PER}sonalization), a dual-track prefix framework for scalable LLM personalization. Recommendation systems have long shown that user preferences are multi-timescale, with persistent interests and transient, context-dependent intents~\cite{zhou2018deep,sun2020go}, and that temporal dynamics are crucial for sequential user modeling~\cite{li2020time}.
As illustrated in Figure~\ref{fig:challenge}, TAP-PER follows this principle by decomposing personalization into a \emph{user-state prefix} for long-term preferences and a \emph{query-aware record prefix} for dynamically activating relevant historical records. It further incorporates relative time gaps and sequence order into record selection, and uses a shared LoRA-based bridge to integrate the prefix signals end-to-end, eliminating explicit prompt construction while keeping per-user parameters lightweight.
In summary, our contributions are as follows:
\begin{itemize}
    \setlength{\itemsep}{0pt}
    \setlength{\topsep}{0pt}
    \setlength{\parsep}{0pt}

    \item
    To address the retrieval sensitivity of prompt-based personalization and the scalability limitations of per-user model adaptation, we propose TAP-PER, a prefix-based framework that represents user preferences as compact learnable states. Instead of serializing user histories into prompts or maintaining heavy per-user adapters, TAP-PER shifts personalization toward end-to-end optimized preference representations.

    \item
    We design a personalized prefix mechanism that decomposes user preferences into two complementary signals: a user-state prefix for persistent preferences and a query-aware record prefix for instance-specific historical evidence.
    The record prefix uses temporal attention with learned time-gap and order-gap biases to capture evolving user interests, while a shared bridge LoRA integrates the personalized prefix signals into the task-adapted backbone.

    \item
    Across six LaMP tasks, TAP-PER consistently outperforms both prompt-based and model-based baselines while shrinking per-user storage by $130\times$ relative to OPPU and cutting the total parameter footprint to roughly half of PER-PCS at the 1{,}000-user scale; its compact per-user prefix state additionally supports lightweight online adaptation from streaming feedback without retraining the shared backbone.
\end{itemize}

\section{Method}

\begin{figure*}[t]
    \centering
    \includegraphics[width=\textwidth]{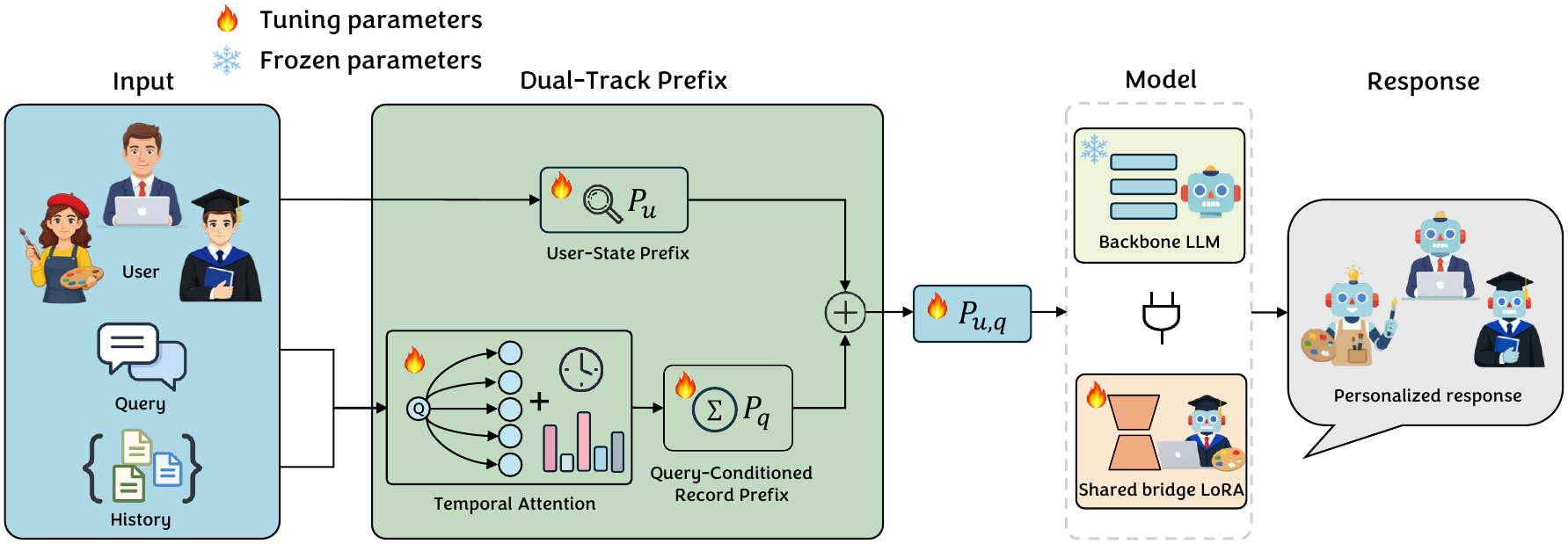}
    \caption{Overview of TAP-PER's personalized prefix learning. With the task-adapted backbone $\theta_{\text{task}}$ frozen, a user-state prefix $\mathbf{P}_u$ is fetched from a per-user embedding table and a query-aware record prefix $\mathbf{P}_q$ is produced by temporal-decayed attention over the user's full interaction history. The two are summed and prepended to the query, while a shared bridge LoRA $\Delta\theta_{\text{bridge}}$ is jointly trained to deeply condition the backbone on the prefix signals. Stage~1 (task-level LoRA adaptation) precedes this stage and is omitted from the diagram.}
    \label{fig:method}
\end{figure*}
\paragraph{Task formulation}
Following prior work~\cite{tan2024democratizing}, we consider a personalized text generation setting over a set of users $\mathcal{U}$. Each user $u \in \mathcal{U}$ is associated with an interaction history $\mathcal{H}_u$. Given an input query $q$, the goal is to generate an output $r$ that reflects the user's preferences.

Formally, we learn a conditional generator:
\begin{equation}
r = \mathrm{LLM}(q, u, \mathcal{H}_u;\, \theta),
\end{equation}
where personalization is achieved by incorporating both user identity and historical interactions.
\paragraph{Overall Framework}
TAP-PER adopts a two-stage training paradigm: \textbf{Stage~1 (Task Adaptation)} fine-tunes the backbone on data aggregated across users to acquire shared task capabilities, and \textbf{Stage~2 (Personalized Prefix Learning)} introduces user-specific prefixes jointly optimized with a shared LoRA-based bridge. Figure~\ref{fig:method} illustrates the Stage~2 pipeline.

\subsection{Task-level Adaptation}

In the first stage, we adapt the backbone model to the target task using data aggregated from all users. Each task is trained independently with its own Stage~1 pipeline. Specifically, we fine-tune the base LLM with Low-Rank Adaptation (LoRA)~\cite{hu2022lora} on query--response pairs $\{(q_i, r_i)\}$, optimizing the cross-entropy loss
\begin{equation}
    \mathcal{L}_{\text{task}} = \sum_i \mathrm{CE}\big(\mathrm{LLM}(q_i; \theta_{\text{task}}), r_i \big),
\end{equation}
where $\theta_{\text{task}}$ denotes the task-adapted parameters (i.e., the base model with LoRA updates), and $\mathrm{CE}(\cdot)$ is the cross-entropy loss. Following LoRA, the weight update for each target matrix is parameterized as $\Delta W = B^{(t)} A^{(t)}$, where $A^{(t)} \in \mathbb{R}^{r \times k}$ and $B^{(t)} \in \mathbb{R}^{d \times r}$ with rank $r \ll \min(d, k)$, while the backbone parameters remain frozen. After optimization, the learned LoRA parameters are merged into the backbone:
\begin{equation}
    W_{\text{task}} = W + B^{(t)} A^{(t)},
\end{equation}
which serves as the initialization for the subsequent personalization stage. Detailed hyperparameter settings are provided in Appendix~\ref{sec:hyperparams}.

\subsection{Personalized Prefix Learning}

In Stage 2, we freeze the task-adapted backbone $\theta_{\text{task}}$ and learn user-specific adaptations through a lightweight prefix-based parameterization, together with a shared bridge LoRA. By merging the Stage~1 LoRA into the backbone before this stage, we decouple task adaptation from personalization: the backbone retains general task capabilities, and the Stage~2 parameters focus exclusively on modeling user preferences without risking catastrophic forgetting of the learned task knowledge.

\paragraph{User-State Prefix.}
Let $L$ denote the prefix length and $d$ the hidden dimension of the backbone. We model persistent user characteristics via a learnable embedding table $\mathbf{E} \in \mathbb{R}^{|\mathcal{U}|\times (L d)}$.
For each user $u$, the embedding $\mathbf{e}_u$ is reshaped into a prefix matrix $\mathbf{P}_u \in \mathbb{R}^{L\times d}$, which captures long-term preferences shared across queries.

\paragraph{Query-Conditioned Record Prefix.}
Given a query $q$ and the user's full interaction history $\mathcal{H}_u=\{h_1,\dots, h_j, \dots, h_n\}$, we first obtain the representation $\mathbf{z}_{h_j}$ for each record $h_j$ by mean-pooling its token embeddings from the frozen backbone's embedding layer. Since the embedding layer is frozen, these record representations are pre-computed once and cached, so that no per-step encoding of history records is required during training or inference. The query representation $\mathbf{z}_q$ is obtained in the same manner. We then compute relevance scores over \emph{all} historical records using a DIN-style attention mechanism~\cite{zhou2018deep}, dispensing with the separate retriever (e.g., BM25) used by RAG-style baselines:
\begin{equation}
s_j = \mathrm{MLP}\!\big(\mathbf{z}_q \| \mathbf{z}_{h_j} \| (\mathbf{z}_q - \mathbf{z}_{h_j}) \| (\mathbf{z}_q \odot \mathbf{z}_{h_j})\big),
\end{equation}
where $\|$ denotes concatenation, $\odot$ denotes element-wise product, and the MLP consists of two linear layers with GELU activation. The scores are then augmented with temporal bias:
\begin{equation}
\tilde{s}_j = s_j - \lambda_t \log(1+\Delta t_j) - \lambda_o \log(1+\Delta\pi_j),
\end{equation}
where $\Delta t_j$ denotes the time gap between the query and record $h_j$, and $\Delta\pi_j$ denotes the order index gap between the query and record $h_j$ in the user history. The decay coefficients $\lambda_t$ and $\lambda_o$ are learnable scalar parameters.

We normalize scores and aggregate historical signals:
\begin{equation}
\alpha_j = \mathrm{softmax}(\tilde{s}_j),
\end{equation}
and construct the query-conditioned prefix as
\begin{equation}
\mathbf{P}_q = \mathrm{MLP}\!\left(\sum_{j=1}^{n} \alpha_j \mathbf{z}_{h_j}\right),
\end{equation}
where the MLP projects the aggregated representation into the prefix space $\mathbb{R}^{L\times d}$. Letting the attention range over the full history rather than a pre-selected top-$k$ subset removes the dependence on an external retriever and lets the model learn end-to-end which records are relevant to the query; we empirically validate this design choice in \S\ref{sec:sensitivity}.

\paragraph{Prefix Composition.}
We combine persistent and query-conditioned signals via element-wise addition:
\begin{equation}
\mathbf{P}_{u,q} = \mathbf{P}_u + \mathbf{P}_q.
\end{equation}
This parameter-free fusion follows the multi-field intuition used in recommender systems and keeps the framework independent of task-specific fusion modules. Appendix~\ref{sec:fusion} compares it with concatenation, gated-sum, and MLP fusion; the more expressive alternatives do not yield consistent gains.

\paragraph{Bridge LoRA.}
Since prefix tokens interact with the backbone only through the attention mechanism, their influence on intermediate representations is limited. To enable deeper integration of personalized signals, we introduce a shared bridge LoRA $\Delta\theta_{\text{bridge}}$, applied to the query, key, value, and output projection layers of each transformer block. Unlike the Stage~1 LoRA, which is merged into the backbone to encode task knowledge, the bridge LoRA is initialized from scratch and trained jointly with the prefix parameters in Stage~2, conditioning the backbone's internal representations on the prefix signals while keeping the number of additional shared parameters small. We empirically validate this design in Section~\ref{sec:ablation}, where removing the bridge causes the largest performance drop. Detailed hyperparameter settings including LoRA ranks are provided in Appendix~\ref{sec:hyperparams}.

\paragraph{Training Objective.}
We optimize the prefix parameters (user embeddings $\mathbf{E}$, the attention MLP, decay coefficients $\lambda_t, \lambda_o$, and the projection MLP) together with the bridge LoRA $\Delta\theta_{\text{bridge}}$:
\begin{equation}
\mathcal{L}_{\text{pers}}
= \sum_{(u,q,r)}
\mathrm{CE} \big(\mathrm{LLM}(\mathbf{P}_{u,q}, q;\, \theta'),\, r\big),
\end{equation}
where $\theta' = \theta_{\text{task}}+\Delta\theta_{\text{bridge}}$.
$\mathbf{P}_u$ and $\mathbf{P}_q$ can be viewed as parameterized counterparts of PAG and RAG, respectively, encoded as end-to-end-optimized prefixes rather than natural-language prompts.

\begin{table*}[t!]
\centering
\small
\resizebox{2\columnwidth}{!}{
\begin{tabular}{llccccccccccccccc}
\toprule
\multicolumn{1}{c}{} & \multicolumn{1}{c}{} & \multicolumn{5}{c}{\textit{Prompt-based}} & \multicolumn{10}{c}{\textit{Model-based}} \\
\cmidrule(lr){3-7}\cmidrule(lr){8-17}
\multicolumn{1}{l}{\textbf{Task}} & \multicolumn{1}{l}{\textbf{Metric}} & \multicolumn{3}{c}{RAG} & \multicolumn{2}{c}{PAG} & \multicolumn{3}{c}{OPPU} & \multicolumn{3}{c}{PER-PCS} & \multicolumn{1}{c}{P2P} & \multicolumn{3}{c}{TAP-PER} \\
\cmidrule(lr){3-5}\cmidrule(lr){6-7}\cmidrule(lr){8-10}\cmidrule(lr){11-13}\cmidrule(lr){14-14}\cmidrule(lr){15-17}
\multicolumn{1}{c}{} & \multicolumn{1}{c}{} & k=1 & k=2 & k=4 & k=0 & k=1 & base & +rag & +pag & base & +rag & +pag & single & base$^\dagger$ & +$\mathbf{P}_q$$^\ddagger$ & +$\mathbf{P}_u$$^\S$ \\ \midrule
LaMP-1: PERSONALIZED   & Acc $\uparrow$ & 0.659 & 0.642 & 0.659 & 0.683 & 0.675 & 0.654 & 0.675 & 0.715 & 0.626 & 0.626 & 0.691 & \underline{0.724} & 0.642 & 0.699 & \textbf{0.732} \\
CITATION IDENTIFICATION& F1 $\uparrow$  & 0.659 & 0.645 & 0.664 & 0.683 & 0.675 & 0.656 & 0.674 & 0.711 & 0.619 & 0.620 & 0.685 & \underline{0.720} & 0.642 & 0.696 & \textbf{0.728} \\ \midrule
LaMP-2N: PERSONALIZED  & Acc $\uparrow$ & 0.790 & 0.796 & 0.803 & 0.787 & 0.790 & 0.794 & 0.809 & \underline{0.817} & 0.748 & 0.798 & 0.809 & 0.803 & 0.789 & 0.808 & \textbf{0.838} \\
NEWS CATEGORIZE        & F1 $\uparrow$  & 0.542 & 0.560 & 0.571 & 0.532 & 0.546 & 0.544 & 0.596 & 0.600 & 0.505 & 0.589 & \underline{0.608} & 0.598 & 0.528 & 0.575 & \textbf{0.622} \\ \midrule
LaMP-2M: PERSONALIZED  & Acc $\uparrow$ & 0.586 & 0.592 & 0.598 & 0.592 & 0.596 & \underline{0.679} & 0.542 & 0.521 & 0.440 & 0.520 & 0.551 & 0.645 & 0.595 & 0.657 & \textbf{0.751} \\
MOVIE TAGGING          & F1 $\uparrow$  & 0.545 & 0.549 & 0.553 & 0.550 & 0.552 & 0.584 & 0.438 & 0.422 & 0.400 & 0.450 & 0.471 & \underline{0.601} & 0.551 & 0.596 & \textbf{0.682} \\ \midrule
LaMP-3: PERSONALIZED   & MAE $\downarrow$ & 0.420 & 0.375 & 0.393 & 0.304 & 0.357 & 0.218 & 0.223 & 0.250 & 0.393 & 0.205 & \underline{0.188} & 0.201 & 0.286 & \underline{0.188} & \textbf{0.170} \\
PRODUCT RATING         & RMSE $\downarrow$& 0.796 & 0.744 & 0.866 & 0.598 & 0.694 & \underline{0.453} & \textbf{0.430} & 0.540 & 0.581 & 0.491 & 0.472 & 0.507 & 0.567 & \underline{0.453} & \textbf{0.430} \\ \midrule
LaMP-4: PERSONALIZED   & R-1 $\uparrow$ & 0.198 & 0.202 & 0.201 & 0.193 & 0.200 & 0.202 & 0.207 & 0.195 & 0.188 & 0.197 & 0.201 & 0.209 & 0.202 & \underline{0.212} & \textbf{0.219} \\
NEWS HEADLINE GEN.     & R-L $\uparrow$ & 0.178 & 0.182 & 0.182 & 0.173 & 0.179 & 0.182 & 0.188 & 0.175 & 0.173 & 0.181 & 0.184 & \underline{0.192} & 0.182 & 0.191 & \textbf{0.198} \\ \midrule
LaMP-5: PERSONALIZED   & R-1 $\uparrow$ & 0.503 & 0.502 & \underline{0.516} & 0.491 & 0.512 & 0.502 & 0.510 & 0.491 & 0.457 & 0.472 & 0.470 & 0.495 & 0.512 & 0.507 & \textbf{0.538} \\
SCHOLARLY TITLE GEN.   & R-L $\uparrow$  & 0.451 & 0.454 & \underline{0.464} & 0.441 & 0.463 & 0.450 & 0.460 & 0.439 & 0.405 & 0.422 & 0.421 & 0.450 & 0.462 & 0.459 & \textbf{0.485} \\ \bottomrule
\end{tabular}
}
\caption{Comparison results on LaMP benchmark across prompt-based and model-based personalization. \textbf{Bold} and \underline{underline} denote the best and second-best result per metric, respectively. $k$ denotes the number of retrieved items, with $k{=}0$ indicating no retrieval. $\uparrow$ indicates higher is better; $\downarrow$ indicates lower is better. For OPPU and PER-PCS, the columns are cumulative: \textit{+rag} adds RAG on top of \textit{base}, and \textit{+pag} further adds PAG on top of \textit{+rag}. For P2P, \textit{single} denotes the single hypernetwork-generated LoRA variant without retrieval or profile stacking. For TAP-PER: $^\dagger$\textit{base} = task-adapted model (Stage~1 only); $^\ddagger$\textit{+$\mathbf{P}_q$} = adding query-aware record prefix; $^\S$\textit{+$\mathbf{P}_u$} = further adding user-state prefix (full model). The $\mathbf{P}_q$ and $\mathbf{P}_u$ columns parallel the \textit{+rag} and \textit{+pag} columns of OPPU/PER-PCS, as they serve analogous roles through learned prefix representations.}
\label{tab:main}
\end{table*}

\section{Experimental Setup}

\paragraph{Datasets.}
Following prior work~\cite{tan2024democratizing,zhang2025proper}, we conduct experiments on the Large Language Model Personalization (LaMP) benchmark~\cite{salemi2024lamp}. LaMP covers a diverse set of personalization tasks spanning classification, regression, and text generation.
We adopt the same data split as in OPPU~\cite{tan2024democratizing}. Detailed task descriptions and dataset statistics are provided in Appendix~\ref{sec:task_desc} and~\ref{sec:data_stats}.

\paragraph{Baselines.}
We compare TAP-PER with two families of personalization baselines: prompt-based methods, including RAG~\cite{salemi2024lamp,lewis2020retrieval} and PAG~\cite{richardson2023integrating}; and model-based methods, including OPPU~\cite{tan2024democratizing}, PER-PCS~\cite{tan2024personalized}, P2P~\cite{tan2025instant}, and PROPER~\cite{zhang2025proper}. P2P encodes a profile built from user history and uses a hypernetwork to generate user-specific LoRA parameters. PROPER progressively combines population-, group-, and user-level adaptation through routed LoRA experts; because adding its results would make the already dense main table unreadable, we report the six-task comparison in Appendix~\ref{sec:proper}.
We use Llama-3.1-8B~\cite{grattafiori2024llama} as the backbone for all methods. BM25~\cite{trotman2014improvements} is used for retrieval-based baselines, whereas TAP-PER directly attends over each user's full history without an external retriever.
More details are provided in Appendix~\ref{sec:baselines} and~\ref{sec:hyperparams}.

\paragraph{Evaluation Metrics.}
Following LaMP~\cite{salemi2024lamp}, we use accuracy and F1-score for classification tasks (LaMP-1, LaMP-2), Mean Absolute Error (MAE) and Root Mean Squared Error (RMSE) for LaMP-3, and ROUGE-1 and ROUGE-L~\cite{lin2004rouge} for text generation tasks (LaMP-4, LaMP-5). Higher values indicate better performance for all metrics, except for MAE and RMSE where lower values are better.

\section{Experimental Results}

We evaluate TAP-PER on the LaMP benchmark through six questions:
\paragraph{RQ1:} Does TAP-PER outperform prompt-based and model-based personalization baselines?
\paragraph{RQ2:} What does each component contribute?
\paragraph{RQ3:} How important are temporal signals?
\paragraph{RQ4:} How scalable is TAP-PER?
\paragraph{RQ5:} How sensitive is TAP-PER to retrieval pre-filtering and prefix length?
\paragraph{RQ6:} Can TAP-PER support online adaptation?

\subsection{Main Results (RQ1)}

Table~\ref{tab:main} compares TAP-PER with prompt-based and model-based personalization baselines. TAP-PER achieves the best performance on all task-metric pairs across the six LaMP tasks.
The gains are especially clear on tasks requiring stronger personalization: compared with the best OPPU variant, TAP-PER improves accuracy by 7.2 points on LaMP-2M and ROUGE-1 by 2.8 points on LaMP-5.
We further verify the same trend with an alternative \texttt{Qwen3-4B} backbone in Appendix~\ref{sec:qwen}.

The results also expose the instability of prompt-based personalization: the best RAG size, the usefulness of PAG, and whether the two should be combined vary substantially across tasks, and this sensitivity persists when RAG/PAG is stacked on OPPU or PER-PCS. By replacing prompt-time context with learned prefixes, TAP-PER avoids this brittleness; the incremental gains from \texttt{base} to $+\mathbf{P}_q$ to $+\mathbf{P}_u$ further indicate that query-aware and user-state modeling are complementary (Section~\ref{sec:ablation}).

\subsection{Component Ablation (RQ2)}
\label{sec:ablation}

To answer RQ2, we ablate the bridge LoRA, the query-aware record prefix $\mathbf{P}_q$, and the user-state prefix $\mathbf{P}_u$ from the full TAP-PER model. As shown in Table~\ref{tab:ablation}, removing any component degrades or does not improve performance, confirming that they provide complementary gains.
The bridge LoRA is important for integrating prefix signals into the task-adapted backbone
(e.g., removing it causes large drops on LaMP-1 and LaMP-5),
while $\mathbf{P}_u$ and $\mathbf{P}_q$ capture different personalization cues:
$\mathbf{P}_u$ is more helpful on tasks requiring stable user preferences
(e.g., LaMP-2M and LaMP-5),
whereas $\mathbf{P}_q$ is especially important when query-specific historical evidence is needed
(e.g., LaMP-3).

\begin{table}[t]
\centering
\scriptsize
\setlength{\tabcolsep}{4pt}
\begin{tabular}{llccccc}
\toprule
\textbf{Task} & \textbf{Metric} & \textbf{Base} & \textbf{Full} & \textbf{$-$bridge} & \textbf{$-\mathbf{P}_q$} & \textbf{$-\mathbf{P}_u$} \\
\midrule
\multirow{2}{*}{LaMP-1}  & Acc $\uparrow$  & 0.642 & \textbf{0.732} & 0.659 & 0.699 & 0.699 \\
                         & F1 $\uparrow$   & 0.642 & \textbf{0.728} & 0.659 & 0.695 & 0.696 \\
\midrule
\multirow{2}{*}{LaMP-2N} & Acc $\uparrow$  & 0.789 & \textbf{0.838} & 0.814 & 0.837 & 0.808 \\
                         & F1 $\uparrow$   & 0.528 & \textbf{0.622} & 0.595 & 0.611 & 0.575 \\
\midrule
\multirow{2}{*}{LaMP-2M} & Acc $\uparrow$  & 0.595 & \textbf{0.751} & 0.702 & 0.733 & 0.657 \\
                         & F1 $\uparrow$   & 0.551 & \textbf{0.682} & 0.624 & 0.659 & 0.596 \\
\midrule
\multirow{2}{*}{LaMP-3}  & MAE $\downarrow$  & 0.286 & \textbf{0.170} & 0.241 & 0.241 & 0.188 \\
                         & RMSE $\downarrow$ & 0.567 & \textbf{0.430} & 0.526 & 0.543 & 0.453 \\
\midrule
\multirow{2}{*}{LaMP-4}  & R-1 $\uparrow$  & 0.202 & \textbf{0.219} & 0.211 & 0.218 & 0.212 \\
                         & R-L $\uparrow$  & 0.182 & \textbf{0.198} & 0.190 & \textbf{0.198} & 0.191 \\
\midrule
\multirow{2}{*}{LaMP-5}  & R-1 $\uparrow$  & 0.512 & \textbf{0.538} & 0.508 & 0.523 & 0.507 \\
                         & R-L $\uparrow$  & 0.462 & \textbf{0.485} & 0.449 & 0.473 & 0.459 \\
\bottomrule
\end{tabular}
\caption{Component ablation of TAP-PER. \textit{Base} is the Stage-1 task-adapted model. \textit{Full} is the complete TAP-PER with all components. \textit{$-$bridge}, \textit{$-\mathbf{P}_q$}, and \textit{$-\mathbf{P}_u$} denote removing the bridge LoRA, query-aware record prefix, and user-state prefix, respectively. \textbf{Bold} denotes the best result per row.}
\label{tab:ablation}
\end{table}

\subsection{Temporal Analysis (RQ3)}
\label{sec:temporal}

To answer RQ3, we examine the effect of the temporal biases in the query-aware record prefix through coefficient ablation and attention visualization.

\paragraph{Temporal-bias ablation.}
We ablate the time-gap and order-gap decay coefficients, $\lambda_t$ and $\lambda_o$, on LaMP-2M and LaMP-5. As shown in Table~\ref{tab:temporal}, the full model performs best across all metrics.
Removing both biases causes the largest drop on LaMP-2M, while disabling either one gives intermediate results, indicating that both temporal recency and sequence order contribute to personalization. The effect is smaller on LaMP-5, suggesting that the usefulness of temporal cues is task-dependent.

\begin{figure*}[t]
    \centering
    \includegraphics[width=\textwidth]{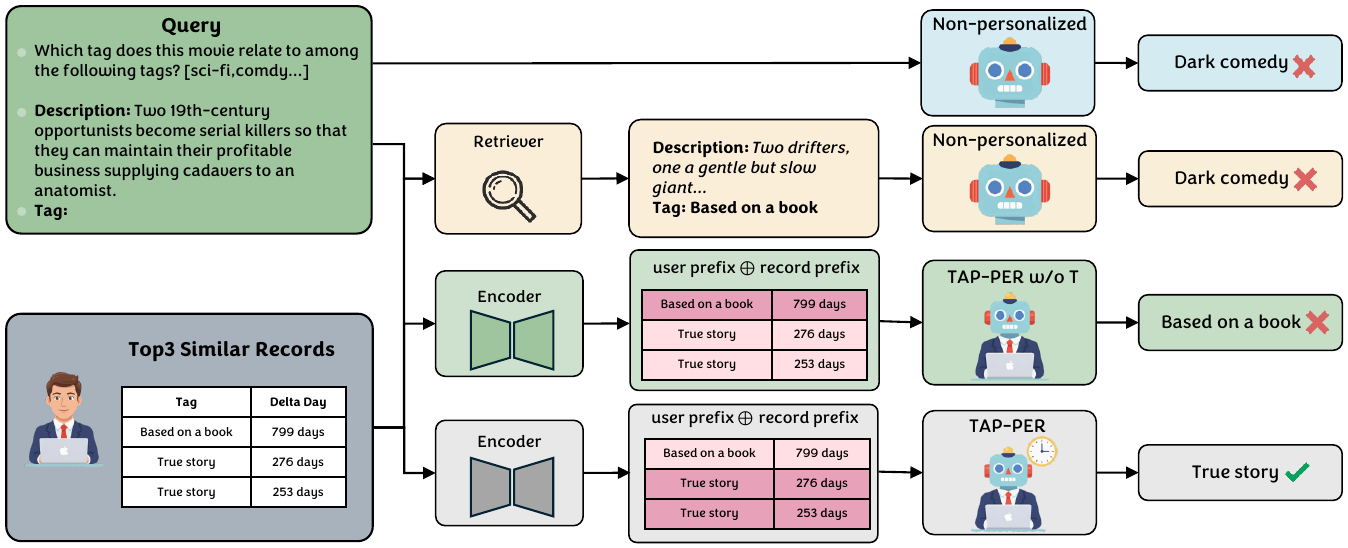}
    \caption{Case study on a representative query from LaMP-2M. The non-personalized baseline and the temporal-free TAP-PER variant predict incorrect tags, whereas the full TAP-PER correctly identifies the ground-truth tag by attending to recent query-relevant records. Darker red indicates higher attention weight.}
    \label{fig:temporal_case}
\end{figure*}

\paragraph{Attention case study.}
Figure~\ref{fig:temporal_case} visualizes a representative LaMP-2M example. Without temporal biases, TAP-PER fails to emphasize recent query-relevant records and predicts an incorrect tag, similar to the non-personalized baseline. In contrast, the full model assigns higher attention to recent relevant histories and recovers the ground-truth tag. This qualitative pattern is consistent with the ablation results in Table~\ref{tab:temporal}.

\begin{table}[t]
\centering
\scriptsize
\setlength{\tabcolsep}{4pt}
\begin{tabular}{llccccc}
\toprule
\textbf{Task} & \textbf{Metric} & \textbf{Base} & \textbf{Full} & $\lambda_t{=}0$ & $\lambda_o{=}0$ & $\lambda_t{=}\lambda_o{=}0$ \\
\midrule
\multirow{2}{*}{LaMP-2M} & Acc $\uparrow$ & 0.595 & \textbf{0.751} & 0.723 & 0.733 & 0.714 \\
                         & F1  $\uparrow$ & 0.551 & \textbf{0.682} & 0.641 & 0.660 & 0.635 \\
\midrule
\multirow{2}{*}{LaMP-5}  & R-1 $\uparrow$ & 0.512 & \textbf{0.538} & 0.516 & 0.511 & 0.513 \\
                         & R-L $\uparrow$ & 0.462 & \textbf{0.485} & 0.464 & 0.452 & 0.464 \\
\bottomrule
\end{tabular}
\caption{Ablation on temporal biases in the query-aware record prefix. \textit{Base} is the Stage-1 task-adapted model. \textit{Full} is the complete TAP-PER. $\lambda_t{=}0$ disables the time-gap decay, $\lambda_o{=}0$ disables the order-gap decay, and $\lambda_t{=}\lambda_o{=}0$ removes both temporal biases while retaining content-based attention. \textbf{Bold} denotes the best result per row.}
\label{tab:temporal}
\end{table}

\subsection{Parameter Efficiency (RQ4)}
\label{sec:efficiency}

\begin{figure}[t]
    \centering
    \includegraphics[width=\columnwidth]{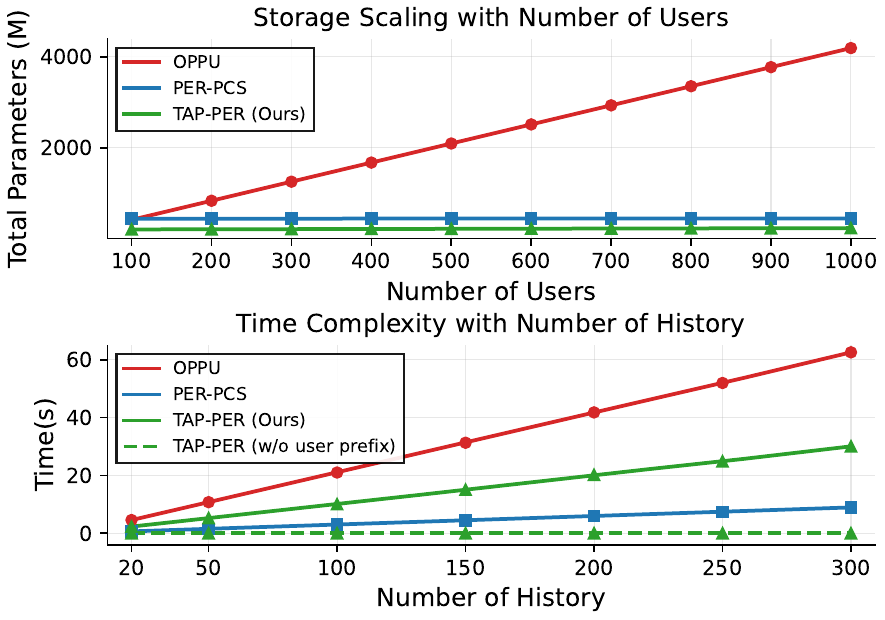}
    \caption{Storage and time scaling comparison. Top: total parameter count as a function of user population. Bottom: per-user training time as a function of history length on LaMP-2M; the dashed green line denotes the TAP-PER variant without user-state prefix $\mathbf{P}_u$, which requires no per-user gradient updates.}
    \label{fig:scaling}
\end{figure}

We evaluate scalability from two aspects: storage cost and per-user training time, comparing TAP-PER with OPPU and PER-PCS in Figure~\ref{fig:scaling}.

\paragraph{Storage efficiency.}
OPPU maintains a dedicated LoRA adapter for each user ($\sim$4.2M parameters), whereas TAP-PER stores only a 32K user-state prefix per user, yielding a $130\times$ reduction in per-user storage. TAP-PER's remaining footprint is \emph{shared} across all users: a bridge LoRA ($\sim$4.2M) plus the DIN-style attention and projection MLPs ($\sim$201M), totaling $\sim$205M shared parameters. As shown in Figure~\ref{fig:scaling} (top), with 1{,}000 users, TAP-PER uses only 5.7\% of OPPU's total parameters and about 53\% of PER-PCS's, showing that it scales mainly through lightweight user embeddings rather than large personalized adapter modules.

This trainable state is separate from the non-trainable history cache. With FP16 record vectors ($d{=}4096$), $H$ cached records require $2Hd$ bytes per user, or approximately 0.8/4.1/8.2\,GB (decimal) in total for 1{,}000 users with $H{=}100/500/1000$. The cache therefore grows linearly with retained history and can be pruned or quantized independently of the trainable prefix.

\paragraph{Training time.}
For each user, TAP-PER only updates the user-state prefix $\mathbf{P}_u$, while the query-aware prefix $\mathbf{P}_q$ is produced by shared parameters. As shown in Figure~\ref{fig:scaling} (bottom), this reduces per-user training time by roughly half compared with OPPU.
PER-PCS is faster because it does not perform per-user gradient updates, but TAP-PER offers a stronger accuracy-efficiency trade-off. In addition, the $\mathbf{P}_q$-only variant requires no per-user training, providing a lightweight option when online user optimization is impractical.

\paragraph{Inference efficiency.}
On one A800 GPU, full-history attention plus prefix projection takes 0.73/0.74/1.04\,ms for 100/500/1000 cached records. For comparison, Llama-3.1-8B has a 108.14\,ms time to first token for a 1{,}024-token prompt and 28.9\,ms/token decoding latency; with a 32-token output, the 1{,}000-record prefix computation is about 0.1\% of total generation time. TAP-PER also avoids the extra prompt tokens introduced by RAG/PAG. Appendix~\ref{sec:generation_efficiency} reports per-user update time on the two generation tasks.

\subsection{Sensitivity Analysis (RQ5)}
\label{sec:sensitivity}

\begin{figure}[t]
    \centering
    \includegraphics[width=\columnwidth]{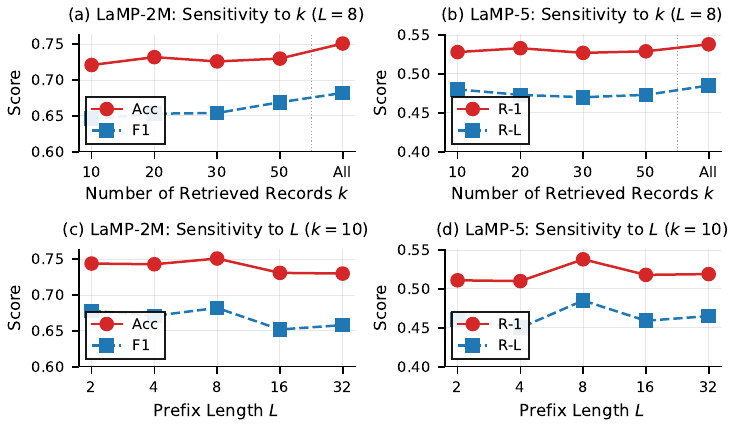}
    \caption{Sensitivity of TAP-PER to the number of retrieved records $k$ (top, with ``All'' denoting our default setting that attends over the full user history) and the prefix length $L$ (bottom). Each subplot shows two metrics: Acc (solid) and F1 (dashed) for LaMP-2M, and R-1 (solid) and R-L (dashed) for LaMP-5.}
    \label{fig:sensitivity}
\end{figure}

To answer RQ5, we study two design choices: history scope and prefix length. For history scope, we insert a BM25 pre-filter before temporal attention and vary the retained records $k \in \{10,20,30,50,\text{All}\}$, where ``All'' denotes our default setting over the full user history. For prefix length, we vary $L \in \{2,4,8,16,32\}$. We evaluate on a representative classification task (LaMP-2M) and a representative generation task (LaMP-5), with results shown in Figure~\ref{fig:sensitivity}.

\paragraph{History scope.}
The ``All'' setting performs best on both tasks. Adding a BM25 pre-filter consistently reduces performance, and varying $k$ within $\{10,20,30,50\}$ does not recover the full-history result. This suggests that explicit pre-retrieval is unnecessary: the temporal attention in $\mathbf{P}_q$ can learn to down-weight less relevant records while preserving useful signals that a sparse retriever may discard.

\paragraph{Prefix length.}
Performance peaks at a moderate prefix length. Short prefixes underfit user-specific signals, while longer prefixes bring little additional gain. We therefore use the full-history setting with $L=8$ as the default configuration.

\subsection{Online Adaptation to Streaming Feedback (RQ6)}
\label{sec:online}

A deployed personalization system should absorb new feedback without full retraining. Since TAP-PER stores user-specific state in a compact $L\times d$ prefix $\mathbf{P}_u$ ($\sim$32K scalars), it can be updated online by optimizing only $\mathbf{P}_u$, leaving all shared parameters unchanged. To answer RQ6, we simulate streaming feedback on LaMP-2M and LaMP-5.

\paragraph{Protocol.}
For each evaluation user, we split the training history chronologically into two equal halves: the earlier 50\% is used for bootstrap (Stage~2) training, and the later 50\% is partitioned into $T{=}5$ equally sized chunks $\{\mathcal{C}_1,\dots,\mathcal{C}_T\}$ that simulate streaming feedback. The original test set is kept unchanged and used for evaluation throughout. After bootstrap training, chunks arrive sequentially; at step $t$, each method updates on $\mathcal{C}_t$ with the same SGD budget and is then evaluated on the fixed test set, yielding a trajectory $R_0,\dots,R_T$.

\paragraph{Variants.}
We compare four variants initialized from the same bootstrap checkpoint: \textbf{Static}, which performs no online update; \textbf{Online-Emb}, which updates only the per-user prefix $\mathbf{P}_u$; \textbf{Online-Full}, which updates $\mathbf{P}_u$ together with the shared $\mathbf{P}_q$ network and bridge LoRA; and \textbf{Full-Retrain}, which retrains TAP-PER from scratch on all data observed so far and serves as an upper bound.

\begin{figure}[t]
    \centering
    \includegraphics[width=\columnwidth]{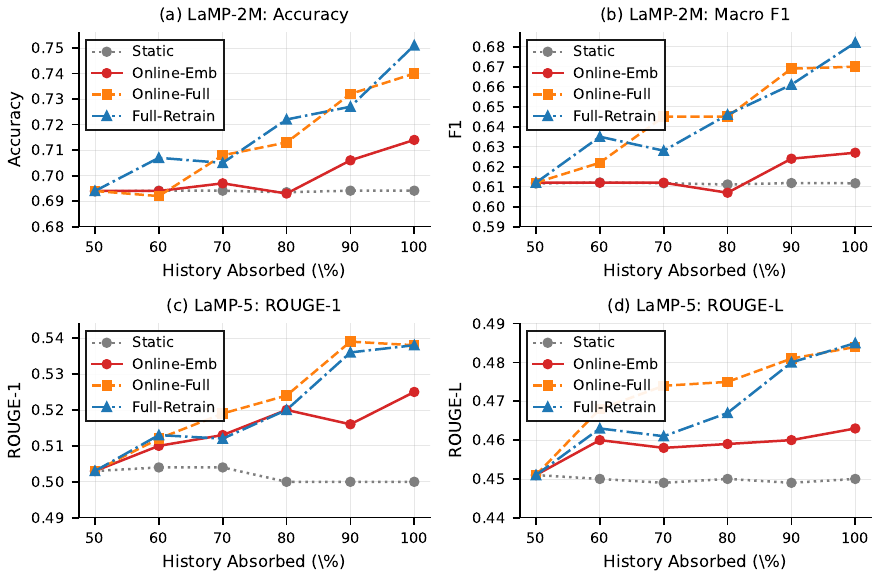}
    \caption{Online adaptation to streaming feedback on LaMP-2M (top) and LaMP-5 (bottom). $x$-axis: fraction of training history absorbed (50\% = bootstrap, 60--100\% = after each of the five streaming chunks); $y$-axis: test-set metric.}
    \label{fig:online}
\end{figure}

\paragraph{Results.}
As shown in Figure~\ref{fig:online}, Static remains flat, confirming that the model cannot benefit from new feedback without online updates. Online-Full nearly matches Full-Retrain at the final step, with less than $\sim$1 point difference on both tasks.
This shows that incremental updates can recover most of the benefit of full retraining without replaying past data.

Online-Emb provides a more lightweight alternative. By updating only the $\sim$32K-parameter user prefix, it steadily improves on LaMP-5, where each user has dense histories, but gains less on LaMP-2M, where each feedback chunk contains only a few samples.
Overall, TAP-PER provides a feasible online-learning pathway: dense user histories can be handled by updating only $\mathbf{P}_u$, while sparse histories may benefit from updating the shared personalization modules, without modifying the backbone or retraining from scratch. This experiment is a controlled probe with chronologically ordered feedback and a fixed test set rather than a full production simulation; robustness to noisy feedback is analyzed in Appendix~\ref{sec:online_robustness}.

\section{Related Work}\label{sec:related_main}

\paragraph{LLM Personalization.}
Existing approaches fall into two families~\cite{liu2025survey,chen2024when,kirk2024benefits}. Prompt-based methods inject user context via retrieval~\cite{salemi2024lamp,mysore-etal-2024-pearl} or natural-language profiles~\cite{richardson2023integrating,li2023teach}, but are sensitive to retrieval and prompt construction. Model-based methods encode preferences into PEFT parameters: OPPU~\cite{tan2024democratizing} trains one adapter per user; PROPER~\cite{zhang2025proper} progressively combines population-, group-, and user-level LoRA experts; PER-PCS~\cite{tan2024personalized} reassembles shared adapter pieces; and P2P~\cite{tan2025instant} generates adapters from profiles via a hypernetwork. HYDRA~\cite{zhuang2024hydra} factorizes reranking and adaptation into shared bases and personal heads for a black-box target LLM. These methods reduce different deployment costs, but still maintain, generate, or assemble personalized parameters and may retain a retrieval/profile pipeline. Complementary methods maintain context outside the backbone: Persona-DB~\cite{sun2025personadb} refines hierarchical personas collaboratively, and LoGo~\cite{wang2025personal} combines local personal memory and collective global memory. Other work optimizes retrieval~\cite{salemi2024optimization}, elicits preferences through dialogue~\cite{wu2025aligning}, or merges preference-specific checkpoints post-hoc~\cite{jang2023personalized}. Earlier UserAdapter~\cite{zhong2021useradapter} attached small per-user modules to pretrained encoders. Closer to our representation-based view, PPlug~\cite{liu2025persona} encodes a user's history into one input-level embedding, while CURP~\cite{wang2026curp} extracts multi-dimensional traits using a user encoder and prototype codebook. TAP-PER instead combines a persistent user-state prefix with a query-aware temporal record prefix, integrated by a shared bridge LoRA.

\paragraph{Prefix and Prompt Tuning.}
Prefix and prompt tuning~\cite{li2021prefix,lester2021power,liu-etal-2022-p} steer frozen transformers via learnable continuous vectors, alongside PEFT variants such as LoRA~\cite{hu2022lora}, adapters~\cite{houlsby2019parameter}, and BitFit~\cite{zaken2022bitfit}. HyperPrompt~\cite{ortiz2020hyperprompt} generates task-conditioned prompts with a hypernetwork. In personalization, SPT~\cite{huang2024selective} learns a bank of soft prompts and a context-conditioned retriever for dialogue. TAP-PER instead treats each user as a learnable prefix and pairs it with a query-aware temporal record prefix.

\paragraph{Sequential Modeling in Recommendation.}
Personalized recommendation has long modeled temporal dynamics and query-aware user interests. Early sequential recommenders encode interaction sequences with recurrent or convolutional networks~\cite{hidasi2016gru4rec,tang2018caser}, and self-attentive successors such as SASRec~\cite{kang2018self} and BERT4Rec~\cite{sun2019bert4rec} capture long-range dependencies within a user's history. DIN~\cite{zhou2018deep} introduces query-conditioned attention over user histories, DIEN~\cite{zhou2019deep} models the evolution of user interests over time, and TiSASRec~\cite{li2020time} incorporates time intervals into self-attention to distinguish recent and distant interactions. More recently, LLMs have been applied directly to recommendation, either as rating predictors or preference reasoners~\cite{dai2023uncovering,kang2023llms,petrov2023generative,wang2024user}. TAP-PER ports these inductive biases to LLM personalization via DIN-style attention with learnable time-gap and order-gap decays, complementing the user-state prefix with context-dependent evidence.

\section{Conclusion}

We presented TAP-PER, a prefix-based framework that reformulates LLM personalization as learned user representation modeling. TAP-PER replaces prompt-serialized user histories and heavy per-user adapters with lightweight prefix states, combining a user-state prefix for persistent preferences, a query-aware temporal record prefix over cached histories, and a shared bridge LoRA for effective integration with the task-adapted backbone.
Our results suggest that effective LLM personalization need not rely on retrieval-time prompt engineering or heavy user-specific adapter modules: a compact, end-to-end-optimized user representation can match or surpass both families while reducing per-user trainable storage by $130\times$ over OPPU and the total parameter footprint to about half of PER-PCS at the 1{,}000-user scale. For an unseen user, the shared query-conditioned prefix provides an immediate fallback, and a user-state prefix can be trained asynchronously once sufficient interactions accumulate.

\section*{Limitations}

TAP-PER has several limitations. First, although $\mathbf{P}_q$ attends over the user's full cached history without an external retriever, cache storage still grows linearly with history length, and attention quality may degrade for extremely long or noisy histories. We have not studied cache pruning, quantization, or hybrid learned retrieval in such regimes.
Second, our evaluation focuses on six LaMP tasks. We exclude LaMP-7 because its unpaired history tweets and rewrite queries do not provide the task-consistent history input--output pairs assumed by our supervised Stage~2; LongLaMP would likewise require adapting its protocol and history construction. Evaluation on these and other benchmarks remains future work.
Third, our experiments cover Llama-3.1-8B and Qwen3-4B only, and we have not verified whether the gains hold at substantially larger backbone sizes. The online study is also a controlled simulation with chronological feedback and a fixed clean test set; despite the added label-noise probe, production feedback can be delayed, biased, non-stationary, or adversarial. Finally, unseen users initially lack $\mathbf{P}_u$. Our shared $\mathbf{P}_q$ fallback performs well for held-out users with histories, but users with no interaction history receive only the task-level model until evidence becomes available, and we do not inductively generate $\mathbf{P}_u$ from a profile.

\section*{Ethical Impact}
TAP-PER is designed to reduce direct exposure of raw histories: user preferences are encoded into continuous prefix embeddings and a shared bridge LoRA rather than injected verbatim into model inputs. This may reduce one surface for prompt-level leakage, but it does not establish a formal privacy guarantee~\cite{carlini2021extracting}. The per-user embeddings and history cache still encode behavioral patterns derived from private histories and should be treated as sensitive artifacts: they require encryption, access control, retention limits, and deletion mechanisms, and users should retain the ability to inspect, update, or delete their personalized representations.

Beyond privacy, personalized LLMs can amplify existing biases or reinforce filter-bubble effects by over-fitting to past behavior, and the temporal decay used in TAP-PER---by design---up-weights recent interactions, which may further narrow the range of content surfaced to a user. We therefore recommend periodic auditing of model outputs across demographic and interest groups, together with transparent controls over the strength of personalization, before deploying frameworks like TAP-PER in user-facing applications. All datasets used in this work are publicly released research benchmarks (LaMP), and no additional user data was collected for this study.

\section*{Acknowledgments}
This work is supported by the ``Pioneer'' and ``Leading Goose'' R\&D Program of Zhejiang (No.~2026C02A1236).

\bibliography{custom}

\begin{thebibliography}{50}
\providecommand{\natexlab}[1]{#1}

\bibitem[{Ben~Zaken et~al.(2022)Ben~Zaken, Goldberg, and Ravfogel}]{zaken2022bitfit}
Elad Ben~Zaken, Yoav Goldberg, and Shauli Ravfogel. 2022.
\newblock \href {https://doi.org/10.18653/v1/2022.acl-short.1} {{B}it{F}it: Simple parameter-efficient fine-tuning for transformer-based masked language-models}.
\newblock In \emph{Proceedings of the 60th Annual Meeting of the Association for Computational Linguistics (Volume 2: Short Papers)}, pages 1--9, Dublin, Ireland. Association for Computational Linguistics.

\bibitem[{Brown et~al.(2020)Brown, Mann, Ryder, Subbiah, Kaplan, Dhariwal, Neelakantan, Shyam, Sastry, Askell, Agarwal, Herbert-Voss, Krueger, Henighan, Child, Ramesh, Ziegler, Wu, Winter, Hesse, Chen, Sigler, Litwin, Gray, Chess, Clark, Berner, McCandlish, Radford, Sutskever, and Amodei}]{brown2020gpt3}
Tom~B. Brown, Benjamin Mann, Nick Ryder, Melanie Subbiah, Jared~D. Kaplan, Prafulla Dhariwal, Arvind Neelakantan, Pranav Shyam, Girish Sastry, Amanda Askell, Sandhini Agarwal, Ariel Herbert-Voss, Gretchen Krueger, Tom Henighan, Rewon Child, Aditya Ramesh, Daniel~M. Ziegler, Jeffrey Wu, Clemens Winter, and 12 others. 2020.
\newblock \href {https://proceedings.neurips.cc/paper/2020/hash/1457c0d6bfcb4967418bfb8ac142f64a-Abstract.html} {Language models are few-shot learners}.
\newblock In \emph{Advances in Neural Information Processing Systems}, volume~33, pages 1877--1901. Curran Associates, Inc.

\bibitem[{Carlini et~al.(2021)Carlini, Tram{\`e}r, Wallace, Jagielski, Herbert-Voss, Lee, Roberts, Brown, Song, Erlingsson, Oprea, and Raffel}]{carlini2021extracting}
Nicholas Carlini, Florian Tram{\`e}r, Eric Wallace, Matthew Jagielski, Ariel Herbert-Voss, Katherine Lee, Adam Roberts, Tom Brown, Dawn Song, {\'U}lfar Erlingsson, Alina Oprea, and Colin Raffel. 2021.
\newblock \href {https://www.usenix.org/conference/usenixsecurity21/presentation/carlini-extracting} {Extracting training data from large language models}.
\newblock In \emph{30th USENIX Security Symposium (USENIX Security 21)}, pages 2633--2650. USENIX Association.

\bibitem[{Chen et~al.(2024)Chen, Liu, Huang, Wu, Liu, Jiang, Pu, Lei, Chen, Wang, Zheng, Lian, and Chen}]{chen2024when}
Jin Chen, Zheng Liu, Xu~Huang, Chenwang Wu, Qi~Liu, Gangwei Jiang, Yuanhao Pu, Yuxuan Lei, Xiaolong Chen, Xingmei Wang, Kai Zheng, Defu Lian, and Enhong Chen. 2024.
\newblock When large language models meet personalization: Perspectives of challenges and opportunities.
\newblock \emph{World Wide Web}, 27(4):42.

\bibitem[{Cui et~al.(2025)Cui, Zhang, Chen, Yuan, Wang, Zuo, Li, Fan, Chen, Chen, Liu, Peng, Bai, Ouyang, Cheng, Zhou, and Ding}]{cui2025entropy}
Ganqu Cui, Yuchen Zhang, Jiacheng Chen, Lifan Yuan, Zhi Wang, Yuxin Zuo, Haozhan Li, Yuchen Fan, Huayu Chen, Weize Chen, Zhiyuan Liu, Hao Peng, Lei Bai, Wanli Ouyang, Yu~Cheng, Bowen Zhou, and Ning Ding. 2025.
\newblock \href {https://doi.org/10.48550/arXiv.2505.22617} {The entropy mechanism of reinforcement learning for reasoning language models}.
\newblock \emph{arXiv preprint arXiv:2505.22617}.

\bibitem[{Dai et~al.(2023)Dai, Shao, Zhao, Yu, Si, Xu, Sun, Zhang, and Xu}]{dai2023uncovering}
Sunhao Dai, Ninglu Shao, Haiyuan Zhao, Weijie Yu, Zihua Si, Chen Xu, Zhongxiang Sun, Xiao Zhang, and Jun Xu. 2023.
\newblock Uncovering {ChatGPT}'s capabilities in recommender systems.
\newblock \emph{arXiv preprint arXiv:2305.02182}.

\bibitem[{Grattafiori et~al.(2024)Grattafiori, Dubey, Jauhri, Pandey, Kadian, Al-Dahle, Letman, Mathur, Schelten, Vaughan, Yang, Fan, Goyal, Hartshorn, Yang, Mitra, Sravankumar, Korenev, Hinsvark, Rao, Zhang, Rodriguez, Gregerson, Spataru, Roziere, Biron, Tang, Chern, Caucheteux, Nayak, Bi, Marra, McConnell, Keller, Touret, Wu, Wong, Ferrer, Nikolaidis, Allonsius, Song, Pintz, Livshits, Wyatt, Esiobu, Choudhary, Mahajan, Garcia-Olano, Perino, Hupkes, Lakomkin, AlBadawy, Lobanova, Dinan, Smith, Radenovic, Guzm{\'a}n, Zhang, Synnaeve, Lee, Anderson, Thattai, Nail, Mialon, Pang, Cucurell, Nguyen, Korevaar, Xu, Touvron, Zarov, Ibarra, Kloumann, Misra, Evtimov, Zhang, Copet, Lee, Geffert, Vranes, Park, Mahadeokar, Shah, van~der Linde, Billock, Hong, Lee, Fu, Chi, Huang, Liu, Wang, Yu, Bitton, Spisak, Park, Rocca, Johnstun, Saxe, Jia, Alwala, Prasad, Upasani, Plawiak, Li, Heafield, Stone, El-Arini, Iyer, Malik, Chiu, Bhalla, Lakhotia, Rantala-Yeary, van~der Maaten, Chen, Tan, Jenkins, Martin, Madaan, Malo, Blecher, Landzaat, de~Oliveira, Muzzi, Pasupuleti, Singh, Paluri, Kardas, Tsimpoukelli, Oldham, Rita, Pavlova, Kambadur, Lewis, Si, Singh, Hassan, Goyal, Torabi, Bashlykov, Bogoychev, Chatterji, Zhang, Duchenne, Çelebi, Alrassy, Zhang, Li, Vasic, Weng, Bhargava, Dubal, Krishnan, Koura, Xu, He, Dong, Srinivasan, Ganapathy, Calderer, Cabral, Stojnic, Raileanu, Maheswari, Girdhar, Patel, Sauvestre, Polidoro, Sumbaly, Taylor, Silva, Hou, Wang, Hosseini, Chennabasappa, Singh, Bell, Kim, Edunov, Nie, Narang, Raparthy, Shen, Wan, Bhosale, Zhang, Vandenhende, Batra, Whitman, Sootla, Collot, Gururangan, Borodinsky, Herman, Fowler, Sheasha, Georgiou, Scialom, Speckbacher, Mihaylov, Xiao, Karn, Goswami, Gupta, Ramanathan, Kerkez, Gonguet, Do, Vogeti, Albiero, Petrovic, Chu, Xiong, Fu, Meers, Martinet, Wang, Wang, Tan, Xia, Xie, Jia, Wang, Goldschlag, Gaur, Babaei, Wen, Song, Zhang, Li, Mao, Coudert, Yan, Chen, Papakipos, Singh, Srivastava, Jain, Kelsey, Shajnfeld, Gangidi, Victoria, Goldstand, Menon, Sharma, Boesenberg, Baevski, Feinstein, Kallet, Sangani, Teo, Yunus, Lupu, Alvarado, Caples, Gu, Ho, Poulton, Ryan, Ramchandani, Dong, Franco, Goyal, Saraf, Chowdhury, Gabriel, Bharambe, Eisenman, Yazdan, James, Maurer, Leonhardi, Huang, Loyd, Paola, Paranjape, Liu, Wu, Ni, Hancock, Wasti, Spence, Stojkovic, Gamido, Montalvo, Parker, Burton, Mejia, Liu, Wang, Kim, Zhou, Hu, Chu, Cai, Tindal, Feichtenhofer, Gao, Civin, Beaty, Kreymer, Li, Adkins, Xu, Testuggine, David, Parikh, Liskovich, Foss, Wang, Le, Holland, Dowling, Jamil, Montgomery, Presani, Hahn, Wood, Le, Brinkman, Arcaute, Dunbar, Smothers, Sun, Kreuk, Tian, Kokkinos, Ozgenel, Caggioni, Kanayet, Seide, Florez, Schwarz, Badeer, Swee, Halpern, Herman, Sizov, Guangyi, Zhang, Lakshminarayanan, Inan, Shojanazeri, Zou, Wang, Zha, Habeeb, Rudolph, Suk, Aspegren, Goldman, Zhan, Damlaj, Molybog, Tufanov, Leontiadis, Veliche, Gat, Weissman, Geboski, Kohli, Lam, Asher, Gaya, Marcus, Tang, Chan, Zhen, Reizenstein, Teboul, Zhong, Jin, Yang, Cummings, Carvill, Shepard, McPhie, Torres, Ginsburg, Wang, Wu, U, Saxena, Khandelwal, Zand, Matosich, Veeraraghavan, Michelena, Li, Jagadeesh, Huang, Chawla, Huang, Chen, Garg, A, Silva, Bell, Zhang, Guo, Yu, Moshkovich, Wehrstedt, Khabsa, Avalani, Bhatt, Mankus, Hasson, Lennie, Reso, Groshev, Naumov, Lathi, Keneally, Liu, Seltzer, Valko, Restrepo, Patel, Vyatskov, Samvelyan, Clark, Macey, Wang, Hermoso, Metanat, Rastegari, Bansal, Santhanam, Parks, White, Bawa, Singhal, Egebo, Usunier, Mehta, Laptev, Dong, Cheng, Chernoguz, Hart, Salpekar, Kalinli, Kent, Parekh, Saab, Balaji, Rittner, Bontrager, Roux, Dollar, Zvyagina, Ratanchandani, Yuvraj, Liang, Alao, Rodriguez, Ayub, Murthy, Nayani, Mitra, Parthasarathy, Li, Hogan, Battey, Wang, Howes, Rinott, Mehta, Siby, Bondu, Datta, Chugh, Hunt, Dhillon, Sidorov, Pan, Mahajan, Verma, Yamamoto, Ramaswamy, Lindsay, Lindsay, Feng, Lin, Zha, Patil, Shankar, Zhang, Zhang, Wang, Agarwal, Sajuyigbe, Chintala, Max, Chen, Kehoe, Satterfield, Govindaprasad, Gupta, Deng, Cho, Virk, Subramanian, Choudhury, Goldman, Remez, Glaser, Best, Koehler, Robinson, Li, Zhang, Matthews, Chou, Shaked, Vontimitta, Ajayi, Montanez, Mohan, Kumar, Mangla, Ionescu, Poenaru, Mihailescu, Ivanov, Li, Wang, Jiang, Bouaziz, Constable, Tang, Wu, Wang, Wu, Gao, Kleinman, Chen, Hu, Jia, Qi, Li, Zhang, Zhang, Adi, Nam, Yu, Wang, Zhao, Hao, Qian, Li, He, Rait, DeVito, Rosnbrick, Wen, Yang, Zhao, and Ma}]{grattafiori2024llama}
Aaron Grattafiori, Abhimanyu Dubey, Abhinav Jauhri, Abhinav Pandey, Abhishek Kadian, Ahmad Al-Dahle, Aiesha Letman, Akhil Mathur, Alan Schelten, Alex Vaughan, Amy Yang, Angela Fan, Anirudh Goyal, Anthony Hartshorn, Aobo Yang, Archi Mitra, Archie Sravankumar, Artem Korenev, Arthur Hinsvark, and 542 others. 2024.
\newblock \href {https://doi.org/10.48550/arXiv.2407.21783} {The {Llama 3} herd of models}.
\newblock \emph{arXiv preprint arXiv:2407.21783}.

\bibitem[{He et~al.(2022)He, Zheng, Tay, Gupta, Du, Aribandi, Zhao, Li, Chen, Metzler, Cheng, and Chi}]{ortiz2020hyperprompt}
Yun He, Steven Zheng, Yi~Tay, Jai Gupta, Yu~Du, Vamsi Aribandi, Zhe Zhao, YaGuang Li, Zhao Chen, Donald Metzler, Heng-Tze Cheng, and Ed~H Chi. 2022.
\newblock \href {https://proceedings.mlr.press/v162/he22f.html} {{HyperPrompt}: Prompt-based task-conditioning of transformers}.
\newblock In \emph{Proceedings of the 39th International Conference on Machine Learning}, volume 162 of \emph{Proceedings of Machine Learning Research}, pages 8678--8690. PMLR.

\bibitem[{Hidasi et~al.(2016)Hidasi, Karatzoglou, Baltrunas, and Tikk}]{hidasi2016gru4rec}
Bal{\'a}zs Hidasi, Alexandros Karatzoglou, Linas Baltrunas, and Domonkos Tikk. 2016.
\newblock Session-based recommendations with recurrent neural networks.
\newblock In \emph{International Conference on Learning Representations}.

\bibitem[{Houlsby et~al.(2019)Houlsby, Giurgiu, Jastrzebski, Morrone, De~Laroussilhe, Gesmundo, Attariyan, and Gelly}]{houlsby2019parameter}
Neil Houlsby, Andrei Giurgiu, Stanislaw Jastrzebski, Bruna Morrone, Quentin De~Laroussilhe, Andrea Gesmundo, Mona Attariyan, and Sylvain Gelly. 2019.
\newblock Parameter-efficient transfer learning for {NLP}.
\newblock In \emph{Proceedings of the 36th International Conference on Machine Learning}, pages 2790--2799.

\bibitem[{Hu et~al.(2022)Hu, Shen, Wallis, Allen-Zhu, Li, Wang, Wang, and Chen}]{hu2022lora}
Edward~J Hu, Yelong Shen, Phillip Wallis, Zeyuan Allen-Zhu, Yuanzhi Li, Shean Wang, Lu~Wang, and Weizhu Chen. 2022.
\newblock \href {https://openreview.net/forum?id=nZeVKeeFYf9} {{LoRA}: Low-rank adaptation of large language models}.
\newblock In \emph{International Conference on Learning Representations}.

\bibitem[{Huang et~al.(2024)Huang, Liu, Ko, Wu, Wang, Zhang, and Tang}]{huang2024selective}
Qiushi Huang, Xubo Liu, Tom Ko, Bo~Wu, Wenwu Wang, Yu~Zhang, and Lilian Tang. 2024.
\newblock \href {https://doi.org/10.18653/v1/2024.findings-acl.959} {Selective prompting tuning for personalized conversations with {LLMs}}.
\newblock In \emph{Findings of the Association for Computational Linguistics: ACL 2024}, pages 16212--16226. Association for Computational Linguistics.

\bibitem[{Jang et~al.(2023)Jang, Kim, Lin, Wang, Hessel, Zettlemoyer, Hajishirzi, Choi, and Ammanabrolu}]{jang2023personalized}
Joel Jang, Seungone Kim, Bill~Yuchen Lin, Yizhong Wang, Jack Hessel, Luke Zettlemoyer, Hannaneh Hajishirzi, Yejin Choi, and Prithviraj Ammanabrolu. 2023.
\newblock \href {https://doi.org/10.48550/arXiv.2310.11564} {Personalized soups: Personalized large language model alignment via post-hoc parameter merging}.
\newblock \emph{arXiv preprint arXiv:2310.11564}.

\bibitem[{Kang and McAuley(2018)}]{kang2018self}
Wang-Cheng Kang and Julian McAuley. 2018.
\newblock Self-attentive sequential recommendation.
\newblock In \emph{2018 IEEE international conference on data mining (ICDM)}, pages 197--206. IEEE.

\bibitem[{Kang et~al.(2023)Kang, Ni, Mehta, Sathiamoorthy, Hong, Chi, and Cheng}]{kang2023llms}
Wang-Cheng Kang, Jianmo Ni, Nikhil Mehta, Maheswaran Sathiamoorthy, Lichan Hong, Ed~Chi, and Derek~Zhiyuan Cheng. 2023.
\newblock Do {LLMs} understand user preferences? evaluating {LLMs} on user rating prediction.
\newblock \emph{arXiv preprint arXiv:2305.06474}.

\bibitem[{Kirk et~al.(2024)Kirk, Vidgen, R{\"o}ttger, and Hale}]{kirk2024benefits}
Hannah~Rose Kirk, Bertie Vidgen, Paul R{\"o}ttger, and Scott~A. Hale. 2024.
\newblock \href {https://doi.org/10.1038/s42256-024-00820-y} {The benefits, risks and bounds of personalizing the alignment of large language models to individuals}.
\newblock \emph{Nature Machine Intelligence}, 6(4):383--392.

\bibitem[{Lester et~al.(2021)Lester, Al-Rfou, and Constant}]{lester2021power}
Brian Lester, Rami Al-Rfou, and Noah Constant. 2021.
\newblock The power of scale for parameter-efficient prompt tuning.
\newblock In \emph{Proceedings of the 2021 Conference on Empirical Methods in Natural Language Processing}, pages 3045--3059.

\bibitem[{Lewis et~al.(2020)Lewis, Perez, Piktus, Petroni, Karpukhin, Goyal, K{\"u}ttler, Lewis, Yih, Rockt{\"a}schel, Riedel, and Kiela}]{lewis2020retrieval}
Patrick Lewis, Ethan Perez, Aleksandra Piktus, Fabio Petroni, Vladimir Karpukhin, Naman Goyal, Heinrich K{\"u}ttler, Mike Lewis, Wen-tau Yih, Tim Rockt{\"a}schel, Sebastian Riedel, and Douwe Kiela. 2020.
\newblock \href {https://proceedings.neurips.cc/paper/2020/hash/6b493230205f780e1bc26945df7481e5-Abstract.html} {Retrieval-augmented generation for knowledge-intensive {NLP} tasks}.
\newblock In \emph{Advances in Neural Information Processing Systems}, volume~33, pages 9459--9474. Curran Associates, Inc.

\bibitem[{Li et~al.(2023)Li, Zhang, Mei, Wang, Hombaiah, Liang, and Bendersky}]{li2023teach}
Cheng Li, Mingyang Zhang, Qiaozhu Mei, Yaqing Wang, Spurthi~Amba Hombaiah, Yi~Liang, and Michael Bendersky. 2023.
\newblock \href {https://doi.org/10.48550/arXiv.2308.07968} {Teach {LLMs} to personalize -- an approach inspired by writing education}.
\newblock \emph{arXiv preprint arXiv:2308.07968}.

\bibitem[{Li et~al.(2020)Li, Wang, and McAuley}]{li2020time}
Jiacheng Li, Yujie Wang, and Julian McAuley. 2020.
\newblock Time interval aware self-attention for sequential recommendation.
\newblock In \emph{Proceedings of the 13th international conference on web search and data mining}, pages 322--330.

\bibitem[{Li and Liang(2021)}]{li2021prefix}
Xiang~Lisa Li and Percy Liang. 2021.
\newblock Prefix-tuning: Optimizing continuous prompts for generation.
\newblock In \emph{Proceedings of the 59th Annual Meeting of the Association for Computational Linguistics and the 11th International Joint Conference on Natural Language Processing (Volume 1: Long Papers)}, pages 4582--4597.

\bibitem[{Lin(2004)}]{lin2004rouge}
Chin-Yew Lin. 2004.
\newblock {ROUGE}: A package for automatic evaluation of summaries.
\newblock In \emph{Text Summarization Branches Out}, pages 74--81.

\bibitem[{Liu et~al.(2025{\natexlab{a}})Liu, Qiu, Li, Dai, Yu, Zhu, Hu, Yang, Chua, and King}]{liu2025survey}
Jiahong Liu, Zexuan Qiu, Zhongyang Li, Quanyu Dai, Wenhao Yu, Jieming Zhu, Minda Hu, Menglin Yang, Tat-Seng Chua, and Irwin King. 2025{\natexlab{a}}.
\newblock \href {https://doi.org/10.48550/arXiv.2502.11528} {A survey of personalized large language models: Progress and future directions}.
\newblock \emph{arXiv preprint arXiv:2502.11528}.

\bibitem[{Liu et~al.(2025{\natexlab{b}})Liu, Zhu, Wang, Wei, Min, Lu, Wang, Yin, and Dou}]{liu2025persona}
Jiongnan Liu, Yutao Zhu, Shuting Wang, Xiaochi Wei, Erxue Min, Yu~Lu, Shuaiqiang Wang, Dawei Yin, and Zhicheng Dou. 2025{\natexlab{b}}.
\newblock {LLM}s + persona-plug = personalized {LLM}s.
\newblock In \emph{Proceedings of the 63rd Annual Meeting of the Association for Computational Linguistics (Volume 1: Long Papers)}, pages 9373--9385.

\bibitem[{Liu et~al.(2022)Liu, Ji, Fu, Tam, Du, Yang, and Tang}]{liu-etal-2022-p}
Xiao Liu, Kaixuan Ji, Yicheng Fu, Weng Tam, Zhengxiao Du, Zhilin Yang, and Jie Tang. 2022.
\newblock P-tuning: Prompt tuning can be comparable to fine-tuning across scales and tasks.
\newblock In \emph{Proceedings of the 60th Annual Meeting of the Association for Computational Linguistics (Volume 2: Short Papers)}, pages 61--68.

\bibitem[{Mysore et~al.(2024)Mysore, Lu, Wan, Yang, Sarrafzadeh, Menezes, Baghaee, Gonzalez, Neville, and Safavi}]{mysore-etal-2024-pearl}
Sheshera Mysore, Zhuoran Lu, Mengting Wan, Longqi Yang, Bahareh Sarrafzadeh, Steve Menezes, Tina Baghaee, Emmanuel~Barajas Gonzalez, Jennifer Neville, and Tara Safavi. 2024.
\newblock \href {https://doi.org/10.18653/v1/2024.customnlp4u-1.16} {Pearl: Personalizing large language model writing assistants with generation-calibrated retrievers}.
\newblock In \emph{Proceedings of the 1st Workshop on Customizable NLP: Progress and Challenges in Customizing NLP for a Domain, Application, Group, or Individual (CustomNLP4U)}, pages 198--219, Miami, Florida, USA. Association for Computational Linguistics.

\bibitem[{Petrov and Macdonald(2023)}]{petrov2023generative}
Aleksandr~V Petrov and Craig Macdonald. 2023.
\newblock Generative sequential recommendation with {GPTRec}.
\newblock In \emph{Gen-IR Workshop at SIGIR}.

\bibitem[{Richardson et~al.(2023)Richardson, Zhang, Gillespie, Kar, Singh, Raeesy, Khan, and Sethy}]{richardson2023integrating}
Chris Richardson, Yao Zhang, Kellen Gillespie, Sudipta Kar, Arshdeep Singh, Zeynab Raeesy, Omar~Zia Khan, and Abhinav Sethy. 2023.
\newblock Integrating summarization and retrieval for enhanced personalization via large language models.
\newblock \emph{arXiv preprint arXiv:2310.20081}.

\bibitem[{Salemi et~al.(2024{\natexlab{a}})Salemi, Kallumadi, and Zamani}]{salemi2024optimization}
Alireza Salemi, Surya Kallumadi, and Hamed Zamani. 2024{\natexlab{a}}.
\newblock Optimization methods for personalizing large language models through retrieval augmentation.
\newblock In \emph{Proceedings of SIGIR}.

\bibitem[{Salemi et~al.(2024{\natexlab{b}})Salemi, Mysore, Bendersky, and Zamani}]{salemi2024lamp}
Alireza Salemi, Sheshera Mysore, Michael Bendersky, and Hamed Zamani. 2024{\natexlab{b}}.
\newblock \href {https://doi.org/10.18653/v1/2024.acl-long.399} {{L}a{MP}: When large language models meet personalization}.
\newblock In \emph{Proceedings of the 62nd Annual Meeting of the Association for Computational Linguistics (Volume 1: Long Papers)}, pages 7370--7392. Association for Computational Linguistics.

\bibitem[{Sun et~al.(2025)Sun, Yang, Reddy, Fung, Chan, Small, Zhai, and Ji}]{sun2025personadb}
Chenkai Sun, Ke~Yang, Revanth~Gangi Reddy, Yi~Ren Fung, Hou~Pong Chan, Kevin Small, ChengXiang Zhai, and Heng Ji. 2025.
\newblock Persona-{DB}: Efficient large language model personalization for response prediction with collaborative data refinement.
\newblock In \emph{Proceedings of the 31st International Conference on Computational Linguistics}, pages 281--296.

\bibitem[{Sun et~al.(2019)Sun, Liu, Wu, Pei, Lin, Ou, and Jiang}]{sun2019bert4rec}
Fei Sun, Jun Liu, Jian Wu, Changhua Pei, Xiao Lin, Wenwu Ou, and Peng Jiang. 2019.
\newblock {BERT4Rec}: Sequential recommendation with bidirectional encoder representations from transformer.
\newblock In \emph{Proceedings of the 28th ACM International Conference on Information and Knowledge Management}, pages 1441--1450.

\bibitem[{Sun et~al.(2020)Sun, Qian, Chen, Liang, Nguyen, and Yin}]{sun2020go}
Ke~Sun, Tieyun Qian, Tong Chen, Yile Liang, Quoc Viet~Hung Nguyen, and Hongzhi Yin. 2020.
\newblock Where to go next: Modeling long-and short-term user preferences for point-of-interest recommendation.
\newblock In \emph{Proceedings of the AAAI conference on artificial intelligence}, volume~34, pages 214--221.

\bibitem[{Tan et~al.(2024{\natexlab{a}})Tan, Liu, and Jiang}]{tan2024personalized}
Zhaoxuan Tan, Zheyuan Liu, and Meng Jiang. 2024{\natexlab{a}}.
\newblock \href {https://doi.org/10.18653/v1/2024.emnlp-main.371} {Personalized pieces: Efficient personalized large language models through collaborative efforts}.
\newblock In \emph{Proceedings of the 2024 Conference on Empirical Methods in Natural Language Processing}, pages 6459--6475. Association for Computational Linguistics.

\bibitem[{Tan et~al.(2024{\natexlab{b}})Tan, Zeng, Tian, Liu, Yin, and Jiang}]{tan2024democratizing}
Zhaoxuan Tan, Qingkai Zeng, Yijun Tian, Zheyuan Liu, Bing Yin, and Meng Jiang. 2024{\natexlab{b}}.
\newblock \href {https://doi.org/10.18653/v1/2024.emnlp-main.372} {Democratizing large language models via personalized parameter-efficient fine-tuning}.
\newblock In \emph{Proceedings of the 2024 Conference on Empirical Methods in Natural Language Processing}, pages 6476--6491. Association for Computational Linguistics.

\bibitem[{Tan et~al.(2026)Tan, Zhang, Wen, Li, Zhang, Chen, Mo, Liu, Zeng, Yin, and Jiang}]{tan2025instant}
Zhaoxuan Tan, Zixuan Zhang, Haoyang Wen, Zheng Li, Rongzhi Zhang, Pei Chen, Fengran Mo, Zheyuan Liu, Qingkai Zeng, Qingyu Yin, and Meng Jiang. 2026.
\newblock \href {https://doi.org/10.18653/v1/2026.acl-long.1081} {Instant personalized large language model adaptation via hypernetwork}.
\newblock In \emph{Proceedings of the 64th Annual Meeting of the Association for Computational Linguistics (Volume 1: Long Papers)}, pages 23557--23580, San Diego, California, United States. Association for Computational Linguistics.

\bibitem[{Tang and Wang(2018)}]{tang2018caser}
Jiaxi Tang and Ke~Wang. 2018.
\newblock Personalized top-n sequential recommendation via convolutional sequence embedding.
\newblock In \emph{Proceedings of WSDM}.

\bibitem[{Trotman et~al.(2014)Trotman, Puurula, and Burgess}]{trotman2014improvements}
Andrew Trotman, Antti Puurula, and Blake Burgess. 2014.
\newblock Improvements to {BM25} and language models examined.
\newblock In \emph{Proceedings of the 2014 Australasian Document Computing Symposium}, pages 58--65.

\bibitem[{Wang et~al.(2026)Wang, Mou, Liu, Huang, and Wei}]{wang2026curp}
Liang Wang, Xinyi Mou, Xiaoyou Liu, Xuanjing Huang, and Zhongyu Wei. 2026.
\newblock {CURP}: Codebook-based continuous user representation for personalized generation with {LLM}s.
\newblock \emph{arXiv preprint arXiv:2602.00742}.

\bibitem[{Wang et~al.(2025)Wang, Wu, Tan, Li, Zhong, Liu, and Zeng}]{wang2025personal}
Zehong Wang, Junlin Wu, ZHaoxuan Tan, Bolian Li, Xianrui Zhong, Zheli Liu, and Qingkai Zeng. 2025.
\newblock From personal to collective: On the role of local and global memory in llm personalization.
\newblock \emph{arXiv preprint arXiv:2509.23767}.

\bibitem[{Wolf et~al.(2020)Wolf, Debut, Sanh, Chaumond, Delangue, Moi, Cistac, Rault, Louf, Funtowicz, Davison, Shleifer, von Platen, Ma, Jernite, Plu, Xu, Le~Scao, Gugger, Drame, Lhoest, and Rush}]{wolf2020transformers}
Thomas Wolf, Lysandre Debut, Victor Sanh, Julien Chaumond, Clement Delangue, Anthony Moi, Pierric Cistac, Tim Rault, R{\'e}mi Louf, Morgan Funtowicz, Joe Davison, Sam Shleifer, Patrick von Platen, Clara Ma, Yacine Jernite, Julien Plu, Canwen Xu, Teven Le~Scao, Sylvain Gugger, and 3 others. 2020.
\newblock \href {https://doi.org/10.18653/v1/2020.emnlp-demos.6} {Transformers: State-of-the-art natural language processing}.
\newblock In \emph{Proceedings of the 2020 Conference on Empirical Methods in Natural Language Processing: System Demonstrations}, pages 38--45, Online. Association for Computational Linguistics.

\bibitem[{Wu et~al.(2024)Wu, Zheng, Qiu, Wang, Gu, Shen, Qin, Zhu, Zhu, Liu, Xiong, and Chen}]{wang2024user}
Likang Wu, Zhi Zheng, Zhaopeng Qiu, Hao Wang, Hongchao Gu, Tingjia Shen, Chuan Qin, Chen Zhu, Hengshu Zhu, Qi~Liu, Hui Xiong, and Enhong Chen. 2024.
\newblock \href {https://doi.org/10.1007/s11280-024-01291-2} {A survey on large language models for recommendation}.
\newblock \emph{World Wide Web}, 27(5):60.

\bibitem[{Wu et~al.(2025)Wu, Fung, Qian, Kim, Hakkani-Tur, and Ji}]{wu2025aligning}
Shujin Wu, Yi~R. Fung, Cheng Qian, Jeonghwan Kim, Dilek Hakkani-Tur, and Heng Ji. 2025.
\newblock Aligning {LLMs} with individual preferences via interaction.
\newblock In \emph{Proceedings of the 31st International Conference on Computational Linguistics}, pages 7648--7662.

\bibitem[{Yang et~al.(2025)Yang, Li, Yang, Zhang, Hui, Zheng, Yu, Gao, Huang, Lv, Zheng, Liu, Zhou, Huang, Hu, Ge, Wei, Lin, Tang, Yang, Tu, Zhang, Yang, Yang, Zhou, Zhou, Lin, Dang, Bao, Yang, Yu, Deng, Li, Xue, Li, Zhang, Wang, Zhu, Men, Gao, Liu, Luo, Li, Tang, Yin, Ren, Wang, Zhang, Ren, Fan, Su, Zhang, Zhang, Wan, Liu, Wang, Cui, Zhang, Zhou, and Qiu}]{yang2025qwen3}
An~Yang, Anfeng Li, Baosong Yang, Beichen Zhang, Binyuan Hui, Bo~Zheng, Bowen Yu, Chang Gao, Chengen Huang, Chenxu Lv, Chujie Zheng, Dayiheng Liu, Fan Zhou, Fei Huang, Feng Hu, Hao Ge, Haoran Wei, Huan Lin, Jialong Tang, and 41 others. 2025.
\newblock \href {https://doi.org/10.48550/arXiv.2505.09388} {{Qwen3} technical report}.
\newblock \emph{arXiv preprint arXiv:2505.09388}.

\bibitem[{Yao et~al.(2025)Yao, Cheng, Wu, Wu, and Tan}]{yao2025diversity}
Jian Yao, Ran Cheng, Xingyu Wu, Jibin Wu, and Kay~Chen Tan. 2025.
\newblock \href {https://doi.org/10.48550/arXiv.2505.23433} {Diversity-aware policy optimization for large language model reasoning}.
\newblock \emph{arXiv preprint arXiv:2505.23433}.

\bibitem[{Zhang et~al.(2025)Zhang, Wu, Zhou, and He}]{zhang2025proper}
Linhai Zhang, Jialong Wu, Deyu Zhou, and Yulan He. 2025.
\newblock \href {https://doi.org/10.18653/v1/2025.acl-long.800} {{PROPER}: A progressive learning framework for personalized large language models with group-level adaptation}.
\newblock In \emph{Proceedings of the 63rd Annual Meeting of the Association for Computational Linguistics (Volume 1: Long Papers)}, pages 16399--16411. Association for Computational Linguistics.

\bibitem[{Zhong et~al.(2021)Zhong, Tang, Wang, Yin, and Duan}]{zhong2021useradapter}
Wanjun Zhong, Duyu Tang, Jiahai Wang, Jian Yin, and Nan Duan. 2021.
\newblock \href {https://doi.org/10.18653/v1/2021.findings-acl.129} {Useradapter: Few-shot user learning in sentiment analysis}.
\newblock In \emph{Findings of the Association for Computational Linguistics: ACL-IJCNLP 2021}, pages 1484--1488. Association for Computational Linguistics.

\bibitem[{Zhou et~al.(2019)Zhou, Mou, Fan, Pi, Bian, Zhou, Zhu, and Gai}]{zhou2019deep}
Guorui Zhou, Na~Mou, Ying Fan, Qi~Pi, Weijie Bian, Chang Zhou, Xiaoqiang Zhu, and Kun Gai. 2019.
\newblock Deep interest evolution network for click-through rate prediction.
\newblock In \emph{Proceedings of the AAAI conference on artificial intelligence}, volume~33, pages 5941--5948.

\bibitem[{Zhou et~al.(2018)Zhou, Zhu, Song, Fan, Zhu, Ma, Yan, Jin, Li, and Gai}]{zhou2018deep}
Guorui Zhou, Xiaoqiang Zhu, Chenru Song, Ying Fan, Han Zhu, Xiao Ma, Yanghui Yan, Junqi Jin, Han Li, and Kun Gai. 2018.
\newblock Deep interest network for click-through rate prediction.
\newblock In \emph{Proceedings of the 24th ACM SIGKDD international conference on knowledge discovery \& data mining}, pages 1059--1068.

\bibitem[{Zhuang et~al.(2024)Zhuang, Sun, Yu, Qiang, Wang, Zhang, and Dai}]{zhuang2024hydra}
Yuchen Zhuang, Haotian Sun, Yue Yu, Rushi Qiang, Qifan Wang, Chao Zhang, and Bo~Dai. 2024.
\newblock \href {https://doi.org/10.52202/079017-3196} {{HYDRA}: Model factorization framework for black-box {LLM} personalization}.
\newblock In \emph{Advances in Neural Information Processing Systems}, volume~37, pages 100783--100815.

\end{thebibliography}

\clearpage
\appendix
\section{Experimental Details}\label{sec:appendix_details}
\subsection{Hyperparameter Settings}\label{sec:hyperparams}

Tables~\ref{tab:hyper_stage1} and~\ref{tab:hyper_stage2} summarize the hyperparameter settings for Stage~1 (task adaptation) and Stage~2 (personalized prefix learning), respectively. Both stages apply LoRA to the query, key, value, and output projection layers. The Stage~1 task LoRA is merged into the backbone after training, while the Stage~2 bridge LoRA remains as a separate shared module.

\begin{table}[t]
\centering
\scriptsize
\begin{tabular}{lcccc}
\toprule
\textbf{Task} & \textbf{Rank} & \textbf{Epochs} & \textbf{learning rate} & \textbf{batch size} \\
\midrule
LaMP-1 & 8 & 3 & 3e\textsuperscript{--4} & 8 \\
LaMP-2N & 8 & 3 & 3e\textsuperscript{--4} & 8 \\
LaMP-2M & 8 & 3 & 3e\textsuperscript{--4} & 8 \\
LaMP-3 & 8 & 3 & 5e\textsuperscript{--4} & 32 \\
LaMP-4 & 8 & 3 & 3e\textsuperscript{--4} & 8 \\
LaMP-5 & 8 & 3 & 3e\textsuperscript{--4} & 8 \\
\bottomrule
\end{tabular}
\caption{Stage~1 (Task Adaptation) hyperparameters.}
\label{tab:hyper_stage1}
\end{table}

\begin{table}[t]
\centering
\scriptsize
\begin{tabular}{lcccc}
\toprule
\textbf{Task} & \textbf{Rank} & \textbf{Epochs} & \textbf{learning rate} & \textbf{batch size} \\
\midrule
LaMP-1 & 4 & 3 & 2e\textsuperscript{--4} & 32 \\
LaMP-2N & 4 & 3 & 1e\textsuperscript{--4} & 16 \\
LaMP-2M & 4 & 3 & 1e\textsuperscript{--4} & 8 \\
LaMP-3 & 4 & 3 & 2e\textsuperscript{--4} & 32 \\
LaMP-4 & 4 & 3 & 2e\textsuperscript{--4} & 32 \\
LaMP-5 & 4 & 3 & 1e\textsuperscript{--4} & 16 \\
\bottomrule
\end{tabular}
\caption{Stage~2 (Personalized Prefix Learning) hyperparameters. The prefix length \textit{L}=8 is fixed across all tasks, and $\mathbf{P}_q$ attends over the user's full history (no retriever).}
\label{tab:hyper_stage2}
\end{table}

\subsection{LaMP Task Descriptions}\label{sec:task_desc}

We employ the LaMP benchmark~\cite{salemi2024lamp}, which encompasses a diverse set of personalization tasks covering both text classification and generation. We evaluate on the following tasks:

\begin{itemize}
    \setlength{\itemsep}{2pt}
    \item \textbf{LaMP-1 (Personalized Citation Identification):} A binary text classification task that involves predicting citation choices. Given a new paper and two candidate references, the model determines which paper will be cited based on the user's historical publications. This task evaluates a model's ability to capture user-specific citation preferences.
    \item \textbf{LaMP-2N (Personalized News Categorization):} A categorical text classification task that involves classifying news articles into one of 15 categories. Given an article written by a user, the model predicts its category using the user's history of articles and their corresponding categories.
    \item \textbf{LaMP-2M (Personalized Movie Tagging):} A categorical text classification task that involves predicting one of 15 tags for a movie based on a user's tagging history. This task evaluates the model's ability to assign tags to a movie description using the user's historical movie-tag pairs.
    \item \textbf{LaMP-3 (Personalized Product Rating):} An ordinal rating prediction task. The model predicts an integer rating from one to five for a given review based on the user's historical review--rating pairs, and is evaluated with MAE and RMSE.
    \item \textbf{LaMP-4 (Personalized News Headline Generation):} A text generation task that involves generating personalized news headlines for given articles based on the author's historical article-title pairs. This task evaluates the model's ability to replicate the author's stylistic nuances in headline generation.
    \item \textbf{LaMP-5 (Personalized Scholarly Title Generation):} A text generation task that involves generating titles for research articles based on the author's historical article-title pairs, extending personalized text generation into scholarly domains.
\end{itemize}

\noindent We exclude the following two tasks from our evaluation:

\begin{itemize}
    \setlength{\itemsep}{2pt}
    \item \textbf{LaMP-6 (Personalized Email Subject Generation):} A text generation task that involves generating subject lines for emails based on the user's historical email-subject pairs. This task is excluded because the corresponding dataset is not publicly available.
    \item \textbf{LaMP-7 (Personalized Tweet Paraphrasing):} A text generation task that rewrites an input tweet in a user's style. Its histories comprise unpaired user tweets, whereas the query requests a rewrite of a separate input. This format does not provide the task-consistent history input--output pairs assumed by our supervised Stage~2 setup, so a faithful extension requires a different training protocol. HYDRA~\cite{zhuang2024hydra} similarly excludes this task. We leave adaptation to LaMP-7 to future work.
\end{itemize}

LongLaMP would offer a useful test of substantially longer histories. However, its data protocol and history construction must be aligned with our supervised Stage~2 setup to avoid an inequivalent comparison. We therefore report this as a limitation rather than retrofit an incompatible protocol, and plan to release the preprocessing and evaluation scripts to facilitate such extensions.

\subsection{Dataset Statistics}\label{sec:data_stats}

Table~\ref{tab:data_stats} summarizes the dataset statistics for each stage. Stage~1 uses all users except the most active ones for task adaptation, while Stage~2 trains on the selected most active users with their full interaction histories. The evaluation set shares the same users as Stage~2.

\begin{table}[t]
\centering
\scriptsize
\setlength{\tabcolsep}{5pt}
\begin{tabular}{lcc|cc|cc}
\toprule
\textbf{Task} & \multicolumn{2}{c|}{\textbf{Stage~1}} & \multicolumn{2}{c|}{\textbf{Stage~2}} & \multicolumn{2}{c}{\textbf{Eval.}} \\
\cmidrule(lr){2-3}\cmidrule(lr){4-5}\cmidrule(lr){6-7}
& \#User & \#Q & \#User & \#History & \#User & \#Q \\
\midrule
LaMP-1  & 6,442  & 7,919  & 100 & 318 & 100 & 123   \\
LaMP-2N & 274    & 3,662  & 49  & 220 & 49  & 6,033 \\
LaMP-2M & 829    & 3,181  & 100 & 56  & 100 & 3,302 \\
LaMP-3  & 19,899 & 22,388 & 101 & 960 & 101 & 112   \\
LaMP-4  & 1,543  & 7,275  & 100 & 270 & 100 & 6,275 \\
LaMP-5  & 14,581 & 16,075 & 101 & 443 & 101 & 107   \\
\bottomrule
\end{tabular}
\caption{Dataset statistics. \#User is the number of users, \#Q is the number of queries, and \#History is the average number of historical records per user for training.}
\label{tab:data_stats}
\end{table}

\subsection{Baseline Details}\label{sec:baselines}

\begin{itemize}
    \setlength{\itemsep}{2pt}
    \item \textbf{RAG (Retrieval-Augmented Generation)~\cite{salemi2024lamp}:} Following the retrieval-augmented personalization method presented in LaMP, the user's query is augmented with the top-$k$ retrieved items from the user's history corpus using BM25. We take $k \in \{1, 2, 4\}$ to study the effect of retrieval size. No model parameters are modified.

    \item \textbf{PAG (Profile-Augmented Generation)~\cite{richardson2023integrating}:} The user's input sequence is concatenated with a natural language profile summarizing the user's preferences and behavior patterns. Following the implementation of \citet{tan2024democratizing}, user profiles are generated using the Vicuna-7B model. We evaluate PAG without retrieval ($k{=}0$) and combined with retrieval augmentation ($k{=}1$), following the setting of \citet{richardson2023integrating}. No model parameters are modified.

    \item \textbf{OPPU (One PEFT Per User)~\cite{tan2024democratizing}:} A dedicated LoRA adapter is trained for each user using their individual data. We evaluate three variants: \textit{base} (LoRA only, no prompt augmentation), \textit{+rag} (LoRA combined with $k{=}1$ retrieved records), and \textit{+pag} (LoRA combined with user profile). LoRA is applied to the same target modules as our method with rank 8.

    \item \textbf{PER-PCS (Personalized Pieces)~\cite{tan2024personalized}:} A collaborative PEFT framework that selects representative ``sharer'' users via clustering, trains individual LoRA adapters and post-hoc gating vectors for each sharer, and assembles personalized PEFT parameters for target users by selecting and weighting sharer pieces in an auto-regressive, training-free manner. We evaluate three variants following OPPU: \textit{base}, \textit{+rag}, and \textit{+pag}.

    \item \textbf{P2P (Profile-to-PEFT)~\cite{tan2025instant}:} A hypernetwork-based personalization framework that maps an encoded user profile directly to a full set of LoRA adapter parameters, eliminating per-user training at deployment time. The user profile is first generated by an upstream LLM summarizing the user's history (following the PAG protocol), then encoded and fed to a hypernetwork that produces a user-specific LoRA. The generated LoRA is loaded into the backbone to produce personalized outputs. We use a single variant per task, since P2P already incorporates PAG (its input is a user profile) and produces a personalized LoRA in one step.

    \item \textbf{PROPER~\cite{zhang2025proper}:} A progressive PEFT framework that bridges population- and user-level adaptation through group-level LoRA experts, a user-aware router, and a LoRA-aware router. The released code depends on a different TRL/Transformers stack, so we implement PROPER in the same Transformers/PEFT pipeline used for all other methods while retaining its progressive adaptation design.
\end{itemize}

\subsection{Comparison with PROPER}\label{sec:proper}

Table~\ref{tab:proper} compares TAP-PER with PROPER under the same backbone, split, training environment, and evaluation protocol. TAP-PER performs better on all six tasks and all reported metrics. This comparison complements Table~\ref{tab:main}: PROPER reduces sparsity through group-level adaptation, whereas TAP-PER stores personalization in compact prefixes and uses a shared temporal history encoder.

\begin{table}[t]
\centering
\scriptsize
\setlength{\tabcolsep}{5pt}
\begin{tabular}{llcc}
\toprule
\textbf{Task} & \textbf{Metric} & \textbf{PROPER} & \textbf{TAP-PER} \\
\midrule
\multirow{2}{*}{LaMP-1}  & Acc $\uparrow$  & 0.691 & \textbf{0.732} \\
                         & F1 $\uparrow$   & 0.695 & \textbf{0.728} \\
\midrule
\multirow{2}{*}{LaMP-2N} & Acc $\uparrow$  & 0.820 & \textbf{0.838} \\
                         & F1 $\uparrow$   & 0.610 & \textbf{0.622} \\
\midrule
\multirow{2}{*}{LaMP-2M} & Acc $\uparrow$  & 0.694 & \textbf{0.751} \\
                         & F1 $\uparrow$   & 0.625 & \textbf{0.682} \\
\midrule
\multirow{2}{*}{LaMP-3}  & MAE $\downarrow$  & 0.206 & \textbf{0.170} \\
                         & RMSE $\downarrow$ & 0.496 & \textbf{0.430} \\
\midrule
\multirow{2}{*}{LaMP-4}  & R-1 $\uparrow$  & 0.203 & \textbf{0.219} \\
                         & R-L $\uparrow$  & 0.188 & \textbf{0.198} \\
\midrule
\multirow{2}{*}{LaMP-5}  & R-1 $\uparrow$  & 0.489 & \textbf{0.538} \\
                         & R-L $\uparrow$  & 0.440 & \textbf{0.485} \\
\bottomrule
\end{tabular}
\caption{Comparison with PROPER on the six LaMP tasks.}
\label{tab:proper}
\end{table}

\subsection{Prefix Fusion Strategies}\label{sec:fusion}

We compare element-wise addition with three alternatives: concatenation followed by projection, a learned gated sum, and an MLP over the concatenated prefixes. Table~\ref{tab:fusion} shows no consistently superior expressive fusion: the MLP is slightly better on LaMP-2M, while addition gives the best LaMP-5 ROUGE-1 and gated-sum gives the best ROUGE-L. We retain addition because it is parameter-free, stable, and competitive, keeping the contribution focused on prefix construction, temporal attention, and bridge integration.

\begin{table}[t]
\centering
\scriptsize
\setlength{\tabcolsep}{4pt}
\begin{tabular}{llcccc}
\toprule
\textbf{Task} & \textbf{Metric} & \textbf{Add} & \textbf{Concat} & \textbf{Gated} & \textbf{MLP} \\
\midrule
\multirow{2}{*}{LaMP-2M} & Acc $\uparrow$ & 0.744 & 0.742 & 0.741 & \textbf{0.747} \\
                         & F1 $\uparrow$  & 0.672 & 0.670 & 0.668 & \textbf{0.674} \\
\midrule
\multirow{2}{*}{LaMP-5}  & R-1 $\uparrow$ & \textbf{0.529} & 0.522 & 0.525 & 0.518 \\
                         & R-L $\uparrow$ & 0.472 & 0.464 & \textbf{0.476} & 0.469 \\
\bottomrule
\end{tabular}
\caption{Comparison of prefix fusion strategies. These controlled runs use the same checkpoint and protocol for all four variants; their absolute values should be compared within this table.}
\label{tab:fusion}
\end{table}

\section{Cold Start and Robustness}\label{sec:cold_start}

\subsection{Unseen-User Generalization}\label{sec:unseen_users}

We conduct a strict user-level split on LaMP-2M and LaMP-5: Stage~2 is trained on 50\% of users and evaluated on the remaining users, who are never seen during personalized training. Since an unseen user has no learned user-state prefix, we set $\mathbf{P}_u=\mathbf{0}$ and use only the shared query-conditioned history prefix $\mathbf{P}_q$. Table~\ref{tab:cold_start} shows that this fallback outperforms Stage~1 and prompt-based baselines on both tasks. Thus, TAP-PER can serve a new user immediately when history is available, without first optimizing user-specific parameters.

\begin{table}[t]
\centering
\scriptsize
\setlength{\tabcolsep}{3.5pt}
\begin{tabular}{llccccc}
\toprule
\textbf{Task} & \textbf{Metric} & \textbf{Stage~1} & \textbf{RAG} & \textbf{PAG} & \textbf{RAG+PAG} & \textbf{$\mathbf{P}_q$ only} \\
\midrule
\multirow{2}{*}{LaMP-2M} & Acc $\uparrow$ & 0.607 & 0.605 & 0.610 & 0.612 & \textbf{0.627} \\
                         & F1 $\uparrow$  & 0.581 & 0.580 & 0.585 & 0.590 & \textbf{0.595} \\
\midrule
\multirow{2}{*}{LaMP-5}  & R-1 $\uparrow$ & 0.528 & 0.522 & 0.514 & 0.518 & \textbf{0.575} \\
                         & R-L $\uparrow$ & 0.480 & 0.484 & 0.464 & 0.485 & \textbf{0.535} \\
\bottomrule
\end{tabular}
\caption{Strict user-level cold-start results. Stage~2 never observes the evaluation users; $\mathbf{P}_u$ is zero and the last column uses only shared $\mathbf{P}_q$.}
\label{tab:cold_start}
\end{table}

\subsection{Early Onboarding with Limited Histories}\label{sec:onboarding}

We further restrict the number of visible histories for Stage~2-unseen users, keep $\mathbf{P}_u=\mathbf{0}$, and perform no per-user update. On LaMP-2M, accuracy remains stable from 0.626 to 0.629 across the tested history budgets. On LaMP-5, ROUGE-L rises from 0.504 with the smallest visible-history budget to 0.540 as more records become available. Updating $\mathbf{P}_u$ from only one or two examples can be unstable; these results support a staged deployment strategy: use shared $\mathbf{P}_q$ during onboarding, then train or update $\mathbf{P}_u$ asynchronously after sufficient reliable interactions accumulate. With no history at all, TAP-PER falls back to the Stage~1 task model.

\subsection{Noisy Streaming Feedback}\label{sec:online_robustness}

To stress-test the controlled online protocol, we randomly corrupt 10--30\% of streaming feedback targets on LaMP-2M and LaMP-5 while evaluating on the original clean test queries. We use the deployment-safe Online-Emb variant, updating only $\mathbf{P}_u$ and freezing all shared components. At 30\% corruption, LaMP-2M F1 remains approximately 0.63; LaMP-5 ROUGE-L changes from 0.476 with clean feedback to 0.468. TAP-PER is therefore reasonably stable under moderate synthetic label noise, but this test does not capture delayed, selection-biased, adversarial, or continuously drifting production feedback.

\section{Additional Efficiency Analysis}\label{sec:additional_efficiency}

\subsection{Generation-Task Update Time}\label{sec:generation_efficiency}

Table~\ref{tab:generation_time} extends the per-user update-time analysis to LaMP-4 and LaMP-5. All measurements use the same A800 environment and report seconds per user. TAP-PER is about 2.5--4$\times$ faster than OPPU as history length grows. PER-PCS remains faster because it avoids per-user gradient updates, whereas full TAP-PER is more accurate in Table~\ref{tab:main} and supports direct updates of a user's compact state.

\begin{table*}[t]
\centering
\scriptsize
\setlength{\tabcolsep}{5pt}
\begin{tabular}{llccccc}
\toprule
\textbf{Task} & \textbf{Method} & $H{=}10$ & $H{=}50$ & $H{=}100$ & $H{=}150$ & $H{=}200$ \\
\midrule
\multirow{3}{*}{LaMP-4} & OPPU    & 4.19 & 19.00 & 37.65 & 54.87 & 76.15 \\
                        & TAP-PER & 1.32 & 5.07  & 9.80  & 14.58 & 19.28 \\
                        & PER-PCS & \textbf{0.57} & \textbf{1.41} & \textbf{2.55} & \textbf{3.57} & \textbf{4.62} \\
\midrule
\multirow{3}{*}{LaMP-5} & OPPU    & 5.12 & 21.80 & 42.87 & 64.20 & 85.41 \\
                        & TAP-PER & 2.08 & 7.68  & 14.36 & 20.98 & 27.85 \\
                        & PER-PCS & \textbf{0.78} & \textbf{2.80} & \textbf{5.49} & \textbf{8.02} & \textbf{10.59} \\
\bottomrule
\end{tabular}
\caption{Per-user update time (seconds) on generation tasks as the number of historical records $H$ increases.}
\label{tab:generation_time}
\end{table*}

\subsection{Generalization to Alternative Backbone}\label{sec:qwen}

To evaluate whether TAP-PER generalizes beyond the primary backbone used in our main experiments (Llama-3.1-8B), we conduct additional experiments using Qwen3-4B~\cite{yang2025qwen3} as an alternative backbone. We compare TAP-PER against the same set of baselines on all six evaluated LaMP tasks. Results are shown in Table~\ref{tab:qwen}.

\begin{table*}[t]
\centering
\small
\resizebox{2\columnwidth}{!}{
\begin{tabular}{llccccccccccccccc}
\toprule
\multicolumn{1}{c}{} & \multicolumn{1}{c}{} & \multicolumn{5}{c}{\textit{Prompt-based}} & \multicolumn{10}{c}{\textit{Model-based}} \\
\cmidrule(lr){3-7}\cmidrule(lr){8-17}
\multicolumn{1}{l}{\textbf{Task}} & \multicolumn{1}{l}{\textbf{Metric}} & \multicolumn{3}{c}{RAG} & \multicolumn{2}{c}{PAG} & \multicolumn{3}{c}{OPPU} & \multicolumn{3}{c}{PER-PCS} & \multicolumn{1}{c}{P2P} & \multicolumn{3}{c}{TAP-PER} \\
\cmidrule(lr){3-5}\cmidrule(lr){6-7}\cmidrule(lr){8-10}\cmidrule(lr){11-13}\cmidrule(lr){14-14}\cmidrule(lr){15-17}
\multicolumn{1}{c}{} & \multicolumn{1}{c}{} & k=1 & k=2 & k=4 & k=0 & k=1 & base & +rag & +pag & base & +rag & +pag & single & base$^\dagger$ & +$\mathbf{P}_q$$^\ddagger$ & +$\mathbf{P}_u$$^\S$ \\ \midrule
LaMP-1: PERSONALIZED   & Acc $\uparrow$ & 0.618 & 0.618 & 0.618 & 0.642 & 0.585 & 0.642 & 0.626 & 0.683 & 0.585 & 0.602 & 0.675 & 0.690 & 0.585 & \underline{0.724} & \textbf{0.813} \\
CITATION IDENTIFICATION& F1 $\uparrow$  & 0.613 & 0.616 & 0.605 & 0.641 & 0.559 & 0.634 & 0.625 & 0.682 & 0.576 & 0.598 & 0.674 & 0.685 & 0.566 & \underline{0.720} & \textbf{0.813} \\ \midrule
LaMP-2N: PERSONALIZED  & Acc $\uparrow$ & 0.770 & 0.777 & 0.787 & 0.783 & 0.785 & 0.785 & 0.793 & \underline{0.805} & 0.662 & 0.749 & 0.768 & 0.787 & 0.763 & 0.785 & \textbf{0.826} \\
NEWS CATEGORIZE        & F1 $\uparrow$  & 0.520 & 0.530 & 0.550 & 0.537 & 0.555 & 0.557 & 0.600 & \underline{0.604} & 0.430 & 0.521 & 0.541 & 0.547 & 0.488 & 0.550 & \textbf{0.609} \\ \midrule
LaMP-2M: PERSONALIZED  & Acc $\uparrow$ & 0.520 & 0.536 & 0.539 & 0.565 & 0.550 & 0.427 & 0.468 & 0.464 & 0.491 & 0.534 & 0.538 & \underline{0.683} & 0.545 & 0.580 & \textbf{0.690} \\
MOVIE TAGGING          & F1 $\uparrow$  & 0.475 & 0.486 & 0.482 & 0.520 & 0.499 & 0.358 & 0.362 & 0.368 & 0.402 & 0.432 & 0.434 & 0.525 & 0.505 & \underline{0.527} & \textbf{0.620} \\ \midrule
LaMP-3: PERSONALIZED   & MAE $\downarrow$ & 0.223 & 0.223 & \underline{0.196} & 0.241 & 0.250 & 0.223 & \underline{0.196} & 0.241 & 0.277 & 0.277 & 0.259 & 0.275 & 0.295 & 0.268 & \textbf{0.195} \\
PRODUCT RATING         & RMSE $\downarrow$& 0.543 & 0.543 & \underline{0.518} & 0.559 & 0.567 & 0.559 & \underline{0.518} & 0.559 & 0.605 & 0.620 & 0.590 & 0.600 & 0.605 & 0.582 & \textbf{0.516} \\ \midrule
LaMP-4: PERSONALIZED   & R-1 $\uparrow$ & 0.176 & 0.175 & \underline{0.184} & 0.174 & 0.179 & 0.163 & 0.168 & 0.165 & 0.148 & 0.157 & 0.156 & \underline{0.184} & 0.174 & 0.175 & \textbf{0.189} \\
NEWS HEADLINE GEN.     & R-L $\uparrow$ & 0.157 & 0.156 & 0.164 & 0.154 & 0.159 & 0.146 & 0.152 & 0.148 & 0.134 & 0.140 & 0.140 & \underline{0.166} & 0.154 & 0.154 & \textbf{0.168} \\ \midrule
LaMP-5: PERSONALIZED   & R-1 $\uparrow$ & 0.489 & \textbf{0.512} & \underline{0.506} & 0.459 & 0.481 & 0.504 & 0.501 & 0.499 & 0.477 & 0.490 & 0.493 & 0.480 & 0.495 & 0.499 & 0.495 \\
SCHOLARLY TITLE GEN.   & R-L $\uparrow$  & 0.422 & \textbf{0.449} & \underline{0.449} & 0.401 & 0.419 & 0.439 & 0.447 & 0.446 & 0.407 & 0.421 & 0.429 & 0.426 & 0.431 & 0.428 & 0.430 \\ \bottomrule
\end{tabular}
}
\caption{Comparison results on LaMP benchmark with Qwen3-4B as backbone. Notation follows Table~\ref{tab:main}.}
\label{tab:qwen}
\end{table*}

The trends largely mirror those on Llama-3.1-8B: TAP-PER (+$\mathbf{P}_u$) is the best on 10 out of 12 task--metric pairs. The exception is LaMP-5, where RAG with $k{=}2$ achieves the best R-1 (0.512) and R-L (0.449), and OPPU also exceeds TAP-PER on both metrics. This suggests that the smaller backbone offers less headroom for prefix-based personalization on short-form scholarly title generation, where direct lexical cues from retrieved abstracts or per-user adaptation remain strong.

\subsection{Prompt Templates}\label{sec:prompts}

Table~\ref{tab:prompts} lists the prompt templates used for each task. For prompt-based methods (RAG, PAG), the user profile and/or retrieved records are prepended before the instruction. For TAP-PER, the prefix representations replace these textual augmentations.

\begin{table*}[!p]
\centering
\footnotesize
\begin{tabular}{p{3.2cm}p{11.5cm}}
\toprule
\textbf{Task} & \textbf{Prompt} \\
\midrule
LaMP-1: Personalized Citation Identification &
\texttt{\{USER PREFIX + RECORD PREFIX\}} / \#\#USER PROFILE: \texttt{\{USER\_PROFILE\}} / \#\#USER HISTORY: \texttt{\{USER\_HISTORY\}} \newline \#\#INSTRUCTION: \newline
Identify the most relevant reference for the listed publication by the researcher. Select the reference paper that is most closely related to the researcher's work. \newline
paper title: \texttt{\{TITLE\}} reference: [1]: ``\texttt{\{OPT1\}}'' [2]: ``\texttt{\{OPT2\}}'' Answer: \\
\midrule
LaMP-2M: Personalized Movie Tagging &
\texttt{\{USER PREFIX + RECORD PREFIX\}} / \#\#USER PROFILE: \texttt{\{USER\_PROFILE\}} / \#\#USER HISTORY: \texttt{\{USER\_HISTORY\}} \newline \#\#INSTRUCTION: \newline
Which tag does this movie relate to among the following tags? Just answer with the tag name without further explanation. tags: [sci-fi, based on a book, comedy, action, twist ending, dystopia, dark comedy, classic, psychology, fantasy, romance, thought-provoking, social commentary, violence, true story] \newline
description: \texttt{\{QUERY\}} tag: \\
\midrule
LaMP-2N: Personalized News Categorization &
\texttt{\{USER PREFIX + RECORD PREFIX\}} / \#\#USER PROFILE: \texttt{\{USER\_PROFILE\}} / \#\#USER HISTORY: \texttt{\{USER\_HISTORY\}} \newline \#\#INSTRUCTION: \newline
Which category does this article relate to among the following categories? Just answer with the category name without further explanation. categories: [travel, education, parents, style \& beauty, entertainment, food \& drink, science \& technology, business, sports, healthy living, women, politics, crime, culture \& arts, religion] \newline
article: \texttt{\{QUERY\}} category: \\
\midrule
LaMP-3: Personalized Product Rating &
\texttt{\{USER PREFIX + RECORD PREFIX\}} / \#\#USER PROFILE: \texttt{\{USER\_PROFILE\}} / \#\#USER HISTORY: \texttt{\{USER\_HISTORY\}} \newline \#\#INSTRUCTION: \newline
What is the score of the following review on a scale of 1 to 5? just answer with 1, 2, 3, 4, or 5 without further explanation. review: \texttt{\{QUERY\}} score: \\
\midrule
LaMP-4: Personalized News Headline Generation &
\texttt{\{USER PREFIX + RECORD PREFIX\}} / \#\#USER PROFILE: \texttt{\{USER\_PROFILE\}} / \#\#USER HISTORY: \texttt{\{USER\_HISTORY\}} \newline \#\#INSTRUCTION: \newline
Generate a headline for the following article. \newline
article: \texttt{\{QUERY\}} headline: \\
\midrule
LaMP-5: Personalized Scholarly Title Generation &
\texttt{\{USER PREFIX + RECORD PREFIX\}} / \#\#USER PROFILE: \texttt{\{USER\_PROFILE\}} / \#\#USER HISTORY: \texttt{\{USER\_HISTORY\}} \newline \#\#INSTRUCTION: \newline
Generate a title for the following abstract of a paper. \newline
abstract: \texttt{\{QUERY\}} title: \\
\bottomrule
\end{tabular}
\caption{Prompt templates for each personalization task. \texttt{\{USER PREFIX + RECORD PREFIX\}} denotes the combined prefix $\mathbf{P}_{u,q} = \mathbf{P}_u + \mathbf{P}_q$, prepended as soft tokens (present only in TAP-PER; absent for all other methods). \texttt{\{USER\_PROFILE\}} is a natural language summary of user preferences (present only in PAG variants). \texttt{\{USER\_HISTORY\}} contains the top-$k$ retrieved records (present only in RAG variants). Each placeholder and its surrounding section header are included only when the corresponding component is active; otherwise the entire block is omitted.}
\label{tab:prompts}
\end{table*}

\subsection{Case Study: Output Comparison}\label{sec:case_outputs}

To complement the quantitative comparison, we provide qualitative output examples. Each table shows the query, the gold answer, and the outputs from a representative non-personalized baseline, prompt-based baselines (RAG and PAG), and two TAP-PER variants (without and with temporal biases). \cmark{} marks a correct answer and \xmark{} an incorrect one.

\begin{table*}[t]
\centering
\footnotesize
\setlength{\tabcolsep}{4pt}

\begin{subtable}{\textwidth}
\centering
\begin{tabular}{p{0.18\textwidth}p{0.78\textwidth}}
\toprule
\textbf{Query (qid 81348)} & As America recovers from the Civil War, one man tries to put the pieces of his life back together but finds himself fighting a new battle on the frontier. Cable is an embittered Confederate soldier who returns from the war to reclaim his Arizona homestead from rebel pioneers who sympathize with the Union war effort. \\
\midrule
\textbf{Gold tag} & based on a book \\
\midrule
\multirow{3}{*}{\textbf{RAG top-3 retrieved}}
 & \textit{based on a book} --- Photographer Robert Kincaid wanders into the life of housewife Francesca Johnson for four days in the 1960s. \\
 & \textit{based on a book} --- Eight people embark on an expedition into the Congo \ldots{} stumble across a race of killer apes. \\
 & \textit{based on a book} --- A vampire relates his epic life story of love, betrayal, loneliness, and dark hunger to an over-curious reporter. \\
\midrule
Non-Personalized & violence~\xmark \\
RAG             & classic~\xmark \\
PAG                        & classic~\xmark \\
TAP-PER (no bias)          & based on a book~\cmark \\
TAP-PER  & based on a book~\cmark \\
\bottomrule
\end{tabular}
\caption{Case study from LaMP-2M.}
\label{tab:case_lamp2m}
\end{subtable}

\vspace{1em}

\begin{subtable}{\textwidth}
\centering
\begin{tabular}{p{0.18\textwidth}p{0.78\textwidth}}
\toprule
\textbf{Abstract} & Implementing security by design in practice often involves the application of threat modeling to elicit security threats and to aid designers in focusing efforts on the most stringent problems first. Existing threat modeling methodologies are capable of generating lots of threats, yet they lack even basic support to triage these threats, except for relying on the expertise and manual assessment by the threat modeler\ldots{} \\
\midrule
\textbf{Gold title} & Risk-based design security analysis. \\
\midrule
\multirow{3}{*}{\textbf{RAG top-3 retrieved}}
 & \textit{Knowledge-enriched security and privacy threat modeling.} --- Creating secure and privacy-protecting systems entails the simultaneous coordination of development activities along three different yet mutually influencing dimensions: translating security and privacy goals to design choices, analyzing the design for threats, and performing a risk analysis of these threats. \\
 & \textit{SPARTA: Security \& Privacy Architecture Through Risk-Driven Threat Assessment.} --- The development of secure and privacy-preserving software systems entails the continuous consideration of the security and privacy aspects of the system under development; the SPARTA prototype facilitates capturing security and privacy design decisions, continuous threat elicitation, and risk analysis. \\
 & \textit{Interaction-Based Privacy Threat Elicitation.} --- Threat modeling involves the systematic identification, elicitation, and analysis of privacy- and/or security-related threats in the context of a specific system; this approach leads to better process guidance and more concrete interpretation of privacy threat types. \\
\midrule
Non-Personalized           & Risk-Based Threat Modeling: A Case Study on SecureDrop \\
RAG                        & Knowledge-enriched security and privacy threat modeling. \\
PAG                        & Risk-Based Threat Modeling: A Case Study on SecureDrop \\
TAP-PER (no bias)          & Risk-Based Threat Modeling: A Case Study on SecureDrop \\
TAP-PER                    & Risk-Based Security Analysis for Threat Modeling \\
\bottomrule
\end{tabular}
\caption{Case study from LaMP-5.}
\label{tab:case_lamp5}
\end{subtable}

\caption{Qualitative output comparison. \cmark{} marks a correct answer and \xmark{} an incorrect one.}
\label{tab:case_studies}
\end{table*}

\subsection{Computation Details}\label{sec:compute}

All experiments are conducted on a single node equipped with 8 NVIDIA A800 80GB GPUs. Training is distributed across GPUs using HuggingFace \texttt{Trainer} with the \texttt{accelerate} backend, launched via \texttt{torchrun} for multi-GPU data parallelism. Stage~1 task adaptation takes approximately 5--30 minutes per task depending on dataset size, and Stage~2 personalized prefix learning takes approximately 10--30 minutes per task. All models are implemented using PyTorch, HuggingFace Transformers~\cite{wolf2020transformers}, and PEFT.

\subsection{Use of Existing Artifacts}\label{sec:artifacts}

Our implementation, processed-data instructions, and reproduction commands are publicly available at \url{https://github.com/heng380/TAP-PER}. All third-party artifacts used in this work are publicly released for research use under their original terms: the LaMP benchmark~\cite{salemi2024lamp} (MIT), Llama-3.1-8B~\cite{grattafiori2024llama} (Llama 3.1 Community License), and Qwen3-4B~\cite{yang2025qwen3} (Apache 2.0). We also consult the software/data supplements released with OPPU~\cite{tan2024democratizing} and PER-PCS~\cite{tan2024personalized}; P2P and PROPER are reproduced in our unified pipeline. Standard libraries (PyTorch, HuggingFace Transformers~\cite{wolf2020transformers}, PEFT) are used under their respective licenses. We use each artifact strictly for its intended research purpose and do not redistribute third-party data or model weights.

\subsection{Potential Future Direction}
Beyond improving personalization accuracy, future work can explore output diversity: a personalized model should not collapse to one narrow pattern when a user may have multiple plausible preferences, styles, or decision criteria. Recent work studies entropy preservation and diversity-aware objectives for language-model reasoning~\cite{cui2025entropy,yao2025diversity}. Adapting such objectives to personalized generation may help preserve multiple high-quality responses while reducing over-personalization.

\subsection{Use of AI Assistants}\label{sec:ai_use}

We used AI assistants in a limited and clearly scoped manner during the preparation of this work. Specifically, AI assistants were used for: (i) code-level autocompletion and refactoring suggestions during the implementation of training and evaluation scripts; (ii) editorial drafting and polishing of author-specified technical content; and (iii) help in formatting LaTeX tables and figures. All research ideas, experimental design, analysis, numerical results, claims, and references were supplied, executed, or independently verified by the authors. All AI-assisted text was reviewed and edited by the authors before inclusion.

\end{document}